\documentclass[12pt,preprintnumbers,floatfix,showpacs,showkeys,tightenlines,nofootinbib]{revtex4}

\usepackage{natbib}
\usepackage{amsmath}
\usepackage{bm}		
\usepackage{graphicx}
\usepackage{dcolumn}	

\def\program#1{{\sc #1}}
\def\defn#1{``#1''}
\def\xdefn#1{#1}

\def\cf{cf.\hbox{}}
\def\ie{i.e.\hbox{}}
\def\eg{eg.\hbox{}}

\def\Cplusplus{\hbox{C\raise.2ex\hbox{\footnotesize ++}}}
\def\P#1{\phantom{#1}}

\def\csmash#1{\hbox to 0em{\hss{#1}\hss}}
\def\xlsy#1#2{{\setbox0=\hbox{#2}\hbox to \wd0{{#1}\hss}}}
\def\xcsy#1#2{{\setbox0=\hbox{#2}\hbox to \wd0{\hss{#1}\hss}}}

\def\ang{{\text{ang}}}

\def\spm{\,{\pm}\,}		
\def\splus{\,{+}\,}		
\def\stimes{\,{\times}\,}	

\def\Jac{\text{\bf J}}

\def\half{\frac{1}{2}}
\def\thalf{{\textstyle \half}}
\def\del{\nabla}
\def\ltsim{\lesssim}

\def\vect#1{\roarrow{#1}}

\def\Dmol{\text{\tt D}}

\def\Mmol{\text{\tt M}}

\def\I{{\text{\sc i}}}
\def\J{{\text{\sc j}}}
\def\K{{\text{\sc k}}}
\def\m{{\text{\tt m}}}

\def\three{{}^{(3)}}

\begin{document}
\title{A Fast Apparent-Horizon Finder for 3-Dimensional Cartesian Grids
       in Numerical Relativity}
\author{Jonathan \surname{Thornburg}}
\email{jthorn@aei.mpg.de}
\homepage{http://www.aei.mpg.de/~jthorn}
\affiliation{Max-Planck-Institut f\"ur Gravitationsphysik,
	     Albert-Einstein-Institut,
	     Am M\"uhlenberg~1, D-14476 Golm, Germany}
\preprint{AEI-2003-049}
%
%


\begin{abstract}
In $3\,{+}\,1$ numerical simulations of dynamic black hole spacetimes,
it's useful to be able to find the apparent horizon(s) (AH) in each slice
of a time evolution.  A number of AH finders are available, but they
often take many minutes to run, so they're too slow to be practically
usable at each time step.  Here I present a new AH finder,
\program{AHFinderDirect}, which is very fast and accurate: at typical
resolutions it takes only a few seconds to find an AH to $\sim 10^{-5} m$
accuracy on a GHz-class processor.

I assume that an AH to be searched for is a Strahlk\"orper
(``star-shaped region'') with respect to some local origin, and so
parameterize the AH shape by $r = h(\text{angle})$ for some single-valued
function $h\,{:}\,\, S^2 \to \Re^+$.  The AH~equation then becomes a nonlinear
elliptic PDE in $h$ on $S^2$, whose coefficients are algebraic functions
of $g_{ij}$, $K_{ij}$, and the Cartesian-coordinate spatial derivatives
of $g_{ij}$.  I discretize $S^2$ using 6~angular patches
(one each in the neighborhood of the $\pm x$, $\pm y$, and $\pm z$ axes)
to avoid coordinate singularities, and finite difference the AH~equation
in the angular coordinates using 4th~order finite differencing.
I solve the resulting system of nonlinear algebraic equations
(for $h$ at the angular grid points) by Newton's method, using a
``symbolic differentiation'' technique to compute the Jacobian matrix.

\program{AHFinderDirect} is implemented as a thorn in the \program{cactus}
computational toolkit, and is freely available by anonymous CVS checkout.
\end{abstract}


\maketitle


\section{Introduction}
\label{sect-introduction}

In $3\,{+}\,1$ numerical relativity, it's often useful to know the positions
and shapes of any black holes in each slice.  These are both key physics
diagnostics, and potentially valuable for helping choose the coordinate
conditions in a numerical evolution.  Moreover, black holes inevitably
contain singularities, which may need to be excised from the computational
domain (\cite{Seidel92a,Anninos94e}).
\footnote{
	 For more recent work on this topic, see (for example)
	 \cite{Cook97a,Alcubierre00a,Brandt00,Alcubierre01a}.
	 }
{}  Since the event horizon can be determined only once the entire future
development of the slice is known,
\footnote{
	 For numerical purposes the usual approximate
	 development to a nearly-stationary state suffices
(\cite{Anninos94f,Libson94a,Caveny-Anderson-Matzner-2003a,Diener03a}).
	 }
{} \ie{} only after a numerical evolution is done, in practice one
usually uses the apparent horizon(s) as a working approximation
which can be computed slice-by-slice while a numerical evolution
is still ongoing.  (Recall that an apparent horizon is always contained
inside an event horizon, and they coincide if the spacetime is stationary
(\cite{Hawking73a}).)  Apparent horizons are also interesting due
to their close relationship to isolated horizons, which have many
useful properties (\cite{Ashtekar98a,Dreyer-etal-2002-isolated-horizons}
and references therein).

There has thus been longstanding interest in algorithms and codes to
find apparent horizons in numerically computed spacetimes (slices).
Here I focus on the case where there are no continuous symmetries
such as axisymmetry, and where the spatial grid is Cartesian.
Many researchers have developed apparent horizon finding
algorithms and codes for this case, for example
\cite{Nakamura84,Tod91,Kemball91a,Baumgarte96,Huq96,Huq00,
Gundlach97a,Anninos98b,Shoemaker-Huq-Matzner-2000,Schnetter02a,Schnetter03a}.
However, with the exception of \cite{Schnetter02a,Schnetter03a},
\footnote{
	 Schnetter (\cite{Schnetter02a,Schnetter03a}) has
	 developed an apparent horizon finding algorithm
	 and code quite similar to mine.  His work and mine
	 were done independently; neither of us learned of
	 the others' work until our own was mostly complete.
	 }
{} the existing numerical codes for apparent horizon finding are generally
very slow, often taking several minutes to find each apparent horizon
even on modern computers.  This is a serious problem, since we would
ideally like to find apparent horizons at each time step of a numerical
evolution, and there may be tens of thousands of such time steps.

In this paper I describe a new numerical apparent horizon algorithm
and code (based on a generalization of the algorithm and code I described
previously for polar-spherical grids (\cite{Thornburg95}))
which is very fast: for typical resolutions it takes only a few seconds
to find an AH, so it's practical to run it at every time step of a
numerical evolution.  This apparent horizon finder is also very accurate,
typically finding apparent horizons to within a few tens of parts
per million in coordinate position, with similar accuracies for
derived quantities such as the apparent horizon area, irreducible mass,
coordinate centroid, etc.
This apparent horizon finder is implemented as a module (\defn{thorn})
\program{AHFinderDirect} in the \program{Cactus} computational toolkit
(\cite{Goodale02a}, \verb|http://www.cactuscode.org|), 
and is freely available (GNU GPL licensed) by anonymous CVS checkout.

In the main body of this paper I give a relatively high-level
description of \program{AHFinderDirect}'s algorithms; in the appendices
I discuss various technical issues in more detail.


\subsection{Notation}

I generally follow the sign and notation conventions of \cite{Wald84}.
In particular, I use the Penrose abstract-index notation, with indices
$i$--$m$ running over the (Cartesian) spatial coordinates
$x^i \equiv (x,y,z)$ in a (spacelike) $3+1$ slice.
$g_{ij}$ is the 3-metric in the slice,
with associated covariant derivative operator $\del_i$.
$K_{ij}$ is the extrinsic curvature of the slice
(I use the sign convention of \cite{York79}, not that of \cite{Wald84})
and $K \equiv K_i{}^i$ is its trace.
$\eta_{ij}$ is the flat 3-metric.
Indices $uvw$ run over generic angular coordinates
$y^u \equiv (\rho,\sigma)$ on the apparent horizon surface.
\defn{$N$-D} abbreviates ``$N$-dimensional''.
In cases where the distinction is important,
I use a prefix $\three$ to denote quantities
defined on a 3-D neighborhood of the apparent horizon surface.

Small-capital indices $\I\J\K$ label angular grid points
on the apparent horizon surface,
and $h[\I]$ is the evaluation of a grid function $h$ at the grid point $\I$.
$\Dmol_u$ and $\Dmol_{uv}$ are finite difference molecules
discretely approximating the angular partial derivatives
$\partial_u$ and $\partial_{uv}$ respectively.
If $\m$ is an index into a finite difference molecule $\Mmol$,
then $\Mmol[\m]$ is an individual molecule coefficient,
and $\m \in \Mmol$ means that this coefficient is nonzero.
Molecule indices may be obtained by subtracting grid point indices
($\m = \J-\I$),
or correspondingly the sum of a grid point index and a molecule index
gives a grid point index ($\J = \I+\m$).


\section{Apparent Horizons}
\label{sect-apparent-horizons}

Given a (spacelike) $3+1$ slice, a \defn{marginally trapped surface}
(MTS) is defined as a closed spacelike 2-surface in the slice, whose
future-pointing outgoing null geodesics have zero expansion $\Theta$.
In terms of the usual $3+1$ variables this condition becomes (\cite{York89})
\begin{equation}
\Theta \equiv
        \del_i n^i + K_{ij} n^i n^j - K = 0
                                                        \label{eqn-Theta(n-i)=0}
\end{equation}
where $n^i$ is the outward-pointing unit normal to the surface.

An \defn{apparent horizon} (AH) is then defined as an outermost MTS
in a slice (there may be multiple MTSs nested inside each other).
In this paper I actually describe an algorithm and code for locating
MTSs, but since the primary application will be the location of AHs,
for convenience of exposition I refer to the MTSs as AHs.

As is common in AH finding, I parameterize the AH surface by first
choosing a local coordinate origin $x_0^i$ inside the AH, then assuming
that the horizon is a \defn{Strahlk\"orper} (\defn{ray body}, or
more commonly \defn{star-shaped region}) about this point.  A
Strahlk\"orper is defined by Minkowski (\cite[p.~108]{Schroeder86})
as
\begin{quote}
a region in $n$-D Euclidean space containing
the origin and whose surface, as seen from the origin,
exhibits only one point in any direction.
\end{quote}

I take $y^u \equiv (\rho,\sigma)$ to be generic angular coordinates
on the AH surface (or equivalently, on the unit 2-sphere $S^2$).
Given these, I then define the AH shape by $r = h(\rho,\sigma)$,
where $r \equiv \left[ \sum_i (x^i - x_0^i)^2 \right]^{1/2}$
is a radial coordinate around the local coordinate origin,
\footnote{
	 Note that I define $r$ to be the {\em flat-space\/}
	 distance from $x_0^i$ to $x^i$ -- there's no use of
	 the 3-metric here.
	 }
{} and the \defn{AH shape function} $h\,{:}\,\, S^2 \to \Re^+$ is
a single-valued function giving the radius of the AH surface as a
function of angular position about the local coordinate origin.


\section{Computing the Expansion $\Theta$ (Continuum)}
\label{sect-computing-Theta-continuum}

To write the expansion~$\Theta$
(and thus the AH~equation~\eqref{eqn-Theta(n-i)=0})
explicitly in terms of this parameterization, \ie{} in terms of $h$'s
1st and 2nd~angular derivatives $\partial_u h$ and $\partial_{uv} h$,
I first define a scalar function which vanishes on the AH surface
and increases outwards, $\three\!F \equiv r - h(\rho,\sigma)$.
I then define a (non-unit) outward-pointing normal covector to the
AH surface as the gradient of this scalar function,
\begin{eqnarray}
s_i \equiv \three\! s_i
	& \equiv &
		\del_i \three\! F
									\\
	& = &	\partial_i \three\! F
			\qquad
			\text{since $F$ is a scalar}
									\\
	& = &	\partial_i r - \partial_i h
									\\
	& = &	\frac{x^i}{r} - X^u_i \partial_u h
			\, \text{,}			
							\label{eqn-s-i(h)}
\end{eqnarray}
where I define the coefficients
$X^u_i \equiv \partial y^u \big/ \partial x^i$.
It's then straightforward to show that
\begin{subequations}
						\label{eqn-partial-i-s-j(h)}
\begin{equation}
\partial_i s_j = \frac{T_{ij}}{r^3}
		 - X^u_{ij} \frac{\partial h}{\partial y^u}
		 - X^u_i X^v_j \frac{\partial^2 h}{\partial y^u \partial y^v}
			\, \text{,}			
\end{equation}
where
\begin{equation}
T_{ij} =
	\begin{cases}
	\displaystyle \sum_{k \ne i} (x^k)^2	& \text{if $i  =  j$}	\\
	\displaystyle - x^i x^j			& \text{if $i \ne j$}	
	\end{cases}
			\, \text{,}			
\end{equation}
and where I also define the coefficients
$X^u_{ij} \equiv \partial^2 y^u \big/ \partial x^i \partial x^j$.
\end{subequations}

The outward-pointing unit normal to the AH surface is then
\begin{equation}
n^i = \frac{s^i}{\|s^k\|}					\\
    = \frac{g^{ij} s_j}{(g^{k\ell} s_k s_\ell)^{1/2}}
			\, \text{,}			
\end{equation}
so the expansion~$\Theta$ is given by
\begin{eqnarray}
\Theta	& \equiv &
		\del_i n^i + K_{ij} n^i n^j - K
									\\
	& = &	\partial_i n^i
		+ (\partial_i \ln \sqrt{g}) n^i
		+ K_{ij} n^i n^j
		- K
									\\
	& = &	\partial_i \frac{g^{ij} s_j}{(g^{k\ell} s_k s_\ell)^{1/2}}
		+ (\partial_i \ln \sqrt{g})
		  \frac{g^{ij} s_j}{(g^{k\ell} s_k s_\ell)^{1/2}}
		+ \frac{K^{ij} s_i s_j}{g^{k\ell} s_k s_\ell}
		- K
									\\
	& = &	  \frac{A}{D^{3/2}}
		+ \frac{B}{D^{1/2}}
		+ \frac{C}{D}
		- K
					\, \text{,}	
							\label{eqn-Theta(ABCD)}
\end{eqnarray}
where
\begin{subequations}
							\label{eqn-ABCD(s-i)}
\begin{eqnarray}
A	& = &	{}
		- (g^{ik} s_k) (g^{j\ell} s_\ell) \partial_i s_j
		- \thalf (g^{ij} s_j) \Bigl[
				      (\partial_i g^{k\ell}) s_k s_\ell
				      \Bigr]
									\\
B	& = &	(\partial_i g^{ij}) s_j
		+ g^{ij} \partial_i s_j
		+ (\partial_i \ln \sqrt{g}) (g^{ij} s_j)
									\\
C	& = &	K^{ij} s_i s_j
									\\
D	& = &	g^{ij} s_i s_j
					\, \text{.}	
\end{eqnarray}
\end{subequations}

Setting $r=h$ in the definitions~\eqref{eqn-s-i(h)}
and~\eqref{eqn-partial-i-s-j(h)} and substituting
into~\eqref{eqn-Theta(ABCD)} and~\eqref{eqn-ABCD(s-i)}
gives $\Theta$ explicitly in terms of $h$ and its 1st and~2nd angular
derivatives,
{} so the AH~equation~\eqref{eqn-Theta(n-i)=0} takes the form
\begin{equation}
\Theta \equiv
	\Theta(h, \partial_u h, \partial_{uv} h;
	       g_{ij}, K_{ij}, \partial_k g_{ij}) = 0
							\label{eqn-Theta(h)=0}
\end{equation}
where the dependence on $g_{ij}$, $K_{ij}$, and $\partial_k g_{ij}$ is
implicit through their position dependence (this is discussed in detail
in section~\ref{sect-computing-Theta-discrete/geometry-interp}).


\section{Solving the Apparent Horizon Equation}
\label{solving-the-AH-eqn}

I view the AH~equation~\eqref{eqn-Theta(h)=0} as an elliptic PDE for
$h$ on $S^2$, and discretize it using standard finite differencing
methods:  I introduce a total of $N_\ang$ angular grid points
$\{ (\rho_\I,\sigma_\I) \}$ on $S^2$, and represent $h$ and $\Theta$
by their values $\{ h_\I \}$ and $\{ \Theta_\I \}$ at these points.
Approximating the angular derivatives $\partial_u h$ and $\partial_{uv} h$
by finite differencing, \eqref{eqn-Theta(h)=0} then becomes a set
of $N_\ang$ nonlinear algebraic equations $\{ \Theta_\I = 0 \}$
for the $N_\ang$ $\{ h_\I \}$ values.

I solve these equations by Newton's method in $N_\ang$ dimensions.
This in turn has several subparts:
\begin{itemize}
\item	The actual Newton's-method iteration algorithm
\item	Computing the (discrete) expansion $\{ \Theta_\I \}$
	given a (discrete) trial AH shape $\{ h_\I \}$
\item	Computing the Jacobian matrix
	$\Jac_{\I\J} \equiv d \Theta_\I \big/ d h_\J$
	given a (discrete) trial AH shape $\{ h_\I \}$
\item	Solving the Newton's-method updating equations
	$\Jac \cdot \delta h = -\Theta$
\end{itemize}
I describe these in detail in the following sections.


\section{Newton's Method}
\label{sect-Newton's-method}

The basic multidimensional Newton's-method algorithm is well known
(see, for example, \cite[section~5.3]{Stoer-Bulirsch-1980}),
but several refinements are necessary for a practical AH finder:

To make Newton's method converge more robustly if the initial guess
is poor, and to limit divergence if the iteration doesn't converge,
\program{AHFinderDirect} limits any single Newton step to have an
$\infty$-norm over the angular grid which is no more than a specified
maximum fraction (10\%~by default) of the mean horizon radius.

Much more sophisticated \defn{modified Newton} algorithms could be
used to achieve faster or more robust convergence
(\eg{} \cite{Bank-Rose-1980,Bank-Rose-1981,
Dennis-Schnabel-1983,MINPACK,Press92,ZIB-TR-90-11,ZIB-TR-91-10}),
but in practice this hasn't been necessary.
\footnote{
	 Additionally, due to the way \program{AHFinderDirect}
	 finds multiple AHs in parallel across multiple processors
	 (discussed in detail in
	  appendix~\ref{app-multiprocessor-parallelization}),
	 it would be difficult to use many of the uniprocessor
	 modified-Newton software packages such as
	 \cite{MINPACK,ZIB-TR-90-11,ZIB-TR-91-10}.
	 }
{}  In particular, the high-spatial-frequency convergence problems
I have previously described for Newton's-method apparent-horizon--finding
(\cite{Thornburg95}), don't seem to occur often in practice.

If the slice doesn't contain an AH (or if either the 3-D Cartesian
grid or the $S^2$ angular grid has insufficient resolution), then
the Newton iteration will probably fail to converge.  In practice,
\program{AHFinderDirect} detects this by limiting the Newton iteration
to a maximum number of iterations.  It's useful to distinguish between
two subcases here:
\begin{itemize}
\item	If we're searching for an AH or AHs at each time step of a
	numerical evolution, and we found this AH at the previous
	time step, then that AH shape probably provides an excellent
	initial guess for this step's Newton iteration, so a relatively
	low maximum-iterations limit is appropriate.  \program{AHFinderDirect}
	uses a default of 10~iterations for this case.
\item	Otherwise (if we do {\em not\/} have a previous--time-step
	AH as an initial guess), in practice the initial guess is likely
	to be rather inaccurate, so a higher maximum-iterations limit
	is appropriate.  \program{AHFinderDirect} uses a default of
	20~iterations for this case.
\end{itemize}

In addition to the maximum-iterations limit, \program{AHFinderDirect}
also aborts the finding of an AH if any trial horizon shape $\{ h_\I \}$
is outside the 3-D Cartesian grid.  Otherwise,
\program{AHFinderDirect} considers an AH to have been found
if and only if the $\infty$-norm of the $\{ \Theta_\I \}$ values
over the angular grid is below a specified threshold ($10^{-8}$ by default).

For better efficiency, in a multiprocessor environment
\program{AHFinderDirect} finds multiple AHs in parallel across
multiple processors.  I describe the algorithm for doing this in
appendix~\ref{app-multiprocessor-parallelization}.


\section{Computing the Expansion $\Theta$ (Discrete)}
\label{sect-computing-Theta-discrete}

Given a trial AH shape $\{ h_\I \}$, I compute the expansion
$\{ \Theta_\I \}$ using~\eqref{eqn-Theta(h)=0}, approximating the
angular derivatives $\partial_u h$ and $\partial_{uv} h$ by the usual
centered 4th~order finite difference molecules $\Dmol_u$ and $\Dmol_{uv}$
respectively.  However, there are several complications in this process,
which I discuss in the following subsections.


\subsection{Geometry Interpolation}
\label{sect-computing-Theta-discrete/geometry-interp}

As shown in section~\ref{sect-computing-Theta-continuum},
the expansion $\Theta$ implicitly depends on the geometry variables
$g_{ij}$, $K_{ij}$, and $\partial_k g_{ij}$ at the AH surface.  In
practice the geometry variables are only known on a (3-D)
Cartesian grid, so they must be interpolated to the AH surface.

Instead of computing the 3-metric derivatives $\partial_k g_{ij}$
on the full 3-D grid and then interpolating these values to the AH
surface, it's much more efficient to do the differentiation only at
the AH-surface points, inside the interpolator:  An interpolator
generally works by (conceptually) locally fitting a fitting function
(usually a low-degree polynomial) to the data points in a neighborhood
of the interpolation point, then evaluating the fitting function at
the interpolation point.  By evaluating the {\em derivative\/} of the
fitting function, the $\partial_k g_{ij}$ values can be interpolated
very cheaply, using only the 3-D input values which are used anyway
for interpolating the $g_{ij}$.

\begin{figure}[tbp]
\begin{center}
\includegraphics{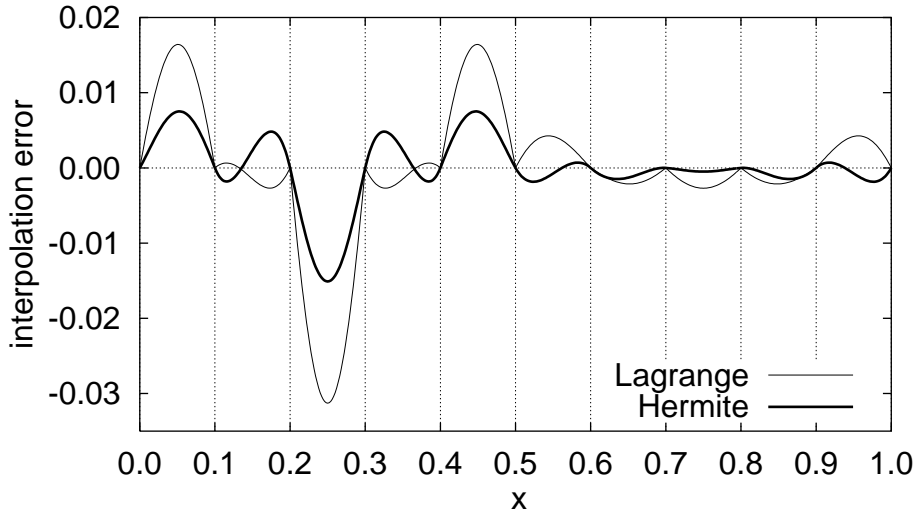}
\end{center}
\caption[Interpolation Errors]
	{
	This figure shows the errors for cubic Lagrange and Hermite
	interpolation of the function $f(x) = \exp[\sin(2 \pi x)]$
	with grid spacing $\Delta x = 0.1$.  Notice that the Lagrange
	error (and hence the Langrange interpolant itself) is
	non-differentiable at the grid points, whereas the
	Hermite error (and interpolant) is differentiable everywhere.
	}
\label{fig-interp-errors}
\end{figure}

Even for $C^\infty$ $g_{ij}$ and $K_{ij}$, the usual Lagrange polynomial
interpolators give results which are continuous, but {\em not differentiable}
(the interpolated $\partial_k g_{ij}$ generically has a jump discontinuity)
at each position where the interpolator switches the set of input 3-D
grid points it uses.  (The non-smoothness of interpolation errors is
discussed in more detail in~\cite[Appendix~F]{Thornburg98}.)
Unfortunately, this lack of smoothness propagates into the AH
equation~\eqref{eqn-Theta(h)=0}, sometimes causing Newton's method
to fail to converge.  To avoid this problem, I use a (cubic) Hermite
interpolator,
{} which guarantees that the interpolated $g_{ij}$ and $K_{ij}$ remain
differentiable, and that the interpolated $\partial_k g_{ij}$ remains
continuous, even when the interpolator switches input--grid-point sets.
Figure~\ref{fig-interp-errors} shows an example of the smoothness
properties of Lagrange and Hermite interpolation, for a simple 1-D
model problem.

While the resulting ($C^0$) smoothness of $\Theta(h)$ isn't quite
ideal for Newton's method, in practice it seems to be sufficient not
to impair convergence, and attaining a higher degree of smoothness
would require a significantly more complicated and expensive
interpolator.


\subsection{Multiple Grid Patches}
\label{sect-computing-Theta-discrete/multiple-grid-patches}

\begin{figure}[tbp]
\begin{center}
\includegraphics[trim=45mm 60mm 20mm 40mm]{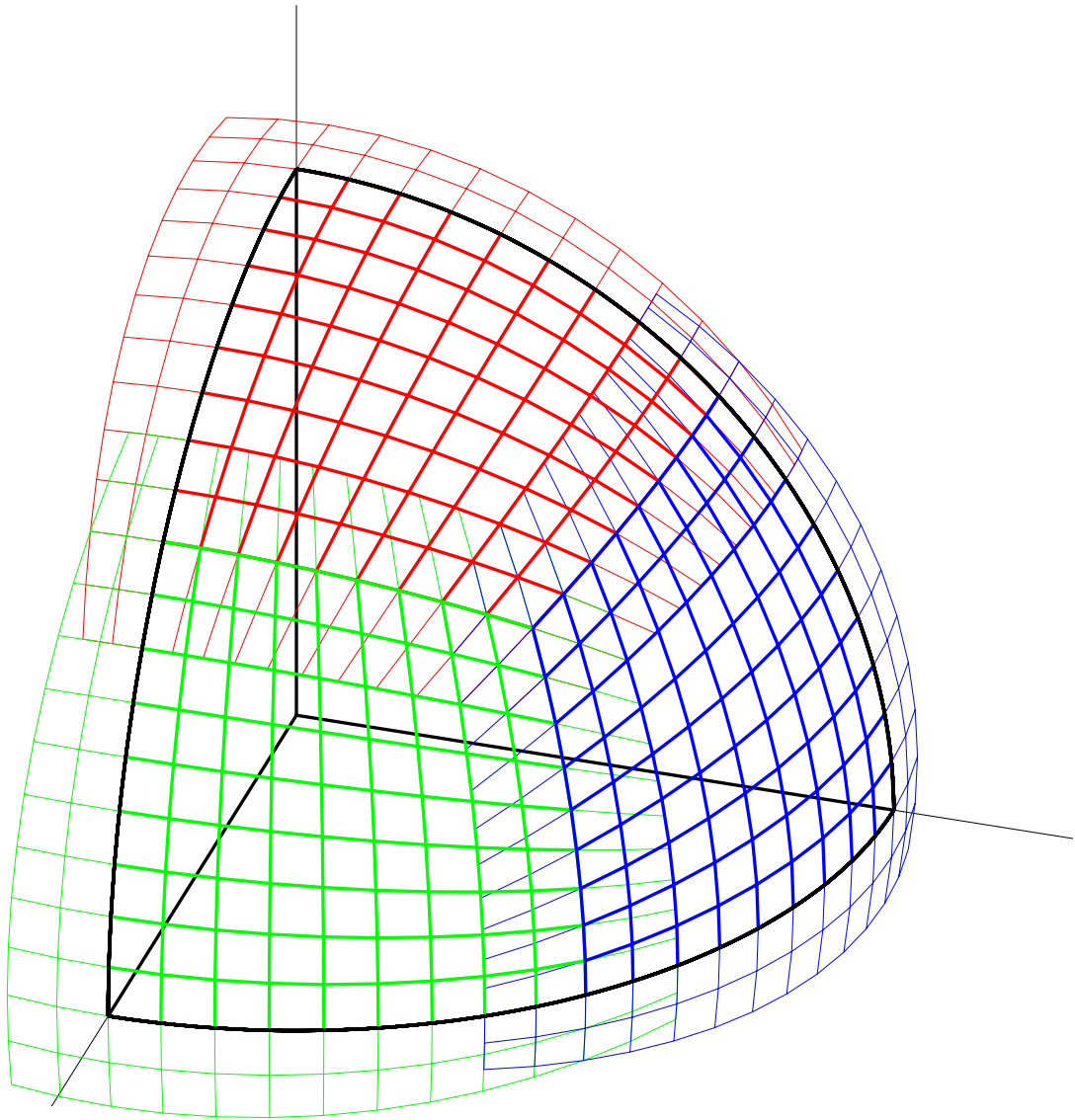}
\end{center}
\caption[Illustration of Multiple Grid Patches]
	{
	This figure shows a multiple-grid-patch system
	covering the $(+,+,+)$~octant of $S^2$ with 3~patches,
	at an angular resolution of $5^\circ$.
	The $+z$, $+x$, and $+y$ patches
	are shown in red, green, and blue respectively.
	Each patch's nominal grid is shown in thick lines;
	the ghost zones are shown in thin lines.
	}
\label{fig-3-patch}
\end{figure}

To avoid $z$~axis coordinate singularities in the angular computations,
I use \xdefn{multiple grid patches} to cover $S^2$.  Figure~\ref{fig-3-patch}
shows an example of this.  In general there are 6~patches, covering
neighborhoods of the $\pm x$, $\pm y$, and $\pm z$ axes respectively.

Each patch's nominal grid (shown in thick lines in figure~\ref{fig-3-patch})
is surrounded by a \defn{ghost zone} (shown in thin lines in
figure~\ref{fig-3-patch}).  Once the $h$ values in the ghost zones
are filled in by symmetry operations and/or interpatch interpolation,
the finite differencing code can ignore the patch boundaries in
computing $\Theta$.  To keep the interpatch interpolation errors
(more precisely, their numerical 2nd~derivatives) from dominating
those of the 4th~order patch-interior angular finite differencing,
I use 5th~order Lagrange polynomial interpolation.  The patch coordinates
$(\rho,\sigma)$ are defined such that adjacent patches always share
a common angular coordinate, so only 1-D interpolation is
required here.  I describe the multiple-patch scheme in detail in
appendix~\ref{app-multiple-patch-details}.

The Jacobian matrix $\Jac_{\I\J}$ must also take into account the
ghost-zone symmetry operations and interpatch interpolations.  This
is conceptually simple, but does require explicitly knowing the Jacobian
(\ie{} the interpolation coefficients) of the interpatch interpolation.
The details are somewhat complicated, and are described in
appendix~\ref{app-multiple-patch-details/Jacobian}.

The multiple-patch scheme works well, but requires a lot of subtle
coding, particularly in handling the ghost-zone updates near patch
corners.  The overall patch infrastructure is currently about
12K~(7K~non-blank non-comment) lines of \Cplusplus{} code, out of a total
of about 25K~(15K) lines of \Cplusplus{} and 2.5K~(1.5K) of Maple in
\program{AHFinderDirect}.
In hindsight, a much simpler scheme might well have sufficed to avoid
$z$~axis problems.  Notably, \cite{Schnetter02a,Schnetter03a} reports
excellent results using a simple latitude-longitude grid on $S^2$, with the
grid points staggered across the north/south poles.  Another possibility
(\cite{Gomez97}) would be to have 2~patches meeting at the equator,
each using stereographic coordinates.


\section{Computing the Jacobian Matrix $\Jac_{\I\J}$}
\label{sect-Jacobian}

If there are $N_\ang$ angular grid points, then the Jacobian matrix
$\Jac_{\I\J} \equiv d \Theta_\I \big/ d h_\J$ is an
$N_\ang \stimes N_\ang$ matrix; $\Jac$ is sparse due to the locality
of the angular finite differencing.  The obvious way to compute $\Jac$
is by numerical perturbation: perturb~$h$ at a single angular grid
point~$\J$, then re-evaluate~$\Theta$\,
\footnote{
	 An important optimization is to only re-evaluate
	 $\Theta$ within an angular-molecule--sized
	 neighborhood of the perturbed point~$\J$.
	 }
{} and determine the $\J$th~column of $\Jac$ from the changes in
$\Theta$.  However, for typical $N_\ang$ values of $300$\,--\,$3000$,
this is very slow (though its relative simplicity makes it useful
for debugging purposes).

Instead of numerical perturbation, \program{AHFinderDirect}
normally uses the \defn{symbolic differentiation} algorithm of
\cite[section~VI]{Thornburg95} to compute $\Jac$ directly from the
angular $\partial_u$ and $\partial_{uv}$ finite difference molecule
coefficients and the (continuum) Jacobian coefficients
$\partial \Theta / \partial (\partial_u h)$ and
$\partial \Theta / \partial (\partial_{uv} h)$.
{}  Temporarily neglecting the interpatch interpolation, the Jacobian
is thus given by
\begin{eqnarray}
\Jac_{\I\J} \equiv \frac{d \Theta_\I}{d h_\J}
	& = &	\left\{
		\begin{tabular}{c@{\quad}l}
		$\displaystyle
		\frac{\partial \Theta}
		     {\partial (\partial_u h)}
		\Dmol_u[\J-\I]$
					& if $\J-\I \in \Dmol_u$    \\[2ex]
		$\displaystyle 0$	& otherwise		    
		\end{tabular}
		\right\}
								\nonumber\\
	&   &	+
		\left\{
		\begin{tabular}{c@{\quad}l}
		$\displaystyle
		\frac{\partial \Theta}
		     {\partial (\partial_{uv} h)}
		\Dmol_{uv}[\J-\I]$
					& if $\J-\I \in \Dmol_{uv}$ \\[2ex]
		$\displaystyle 0$	& otherwise		    
		\end{tabular}
		\right\}
		+
		\left\{
		\begin{tabular}{c@{\quad}l}
		$\displaystyle
		\partial_r \Theta$	& if $\I=\J$		    \\[1ex]
		$\displaystyle 0$	& otherwise		    
		\end{tabular}
		\right\}
							\label{eqn-Jacobian}
\end{eqnarray}
where the first two terms describe the variation in $\Theta$ at a fixed
spatial position with respect to $h$, and the last term describes the
variation in $\Theta$ due to a change in~$h$ changing the evaluation
position of -- and thus the position-dependent coefficients in -- $\Theta$.
Notice that there is no term here for $\partial \Theta \big/ \partial h$,
since this dependence is included in the $\partial_r \Theta$ term.

As mentioned in
section~\ref{sect-computing-Theta-discrete/multiple-grid-patches},
the Jacobian~\eqref{eqn-Jacobian} must be modified to take into acount
the ghost-zone symmetry operations and interpatch interpolations.  This
is described in detail in appendix~\ref{app-multiple-patch-details/Jacobian}.

Because $\Theta$ depends on $g_{ij}$, $K_{ij}$, and $\partial_k g_{ij}$
(\cf~\eqref{eqn-Theta(h)=0}), in theory the $\partial_r \Theta$ term
in~\eqref{eqn-Jacobian} also requires interpolating $\partial_k K_{ij}$
and $\partial_{k\ell} g_{ij}$
(\cf~section~\ref{sect-computing-Theta-discrete/geometry-interp}).
However, doing the computation this way would require a much larger
number of interpolations (a total of 80~geometry-interpolator outputs
instead of~30), and the expressions for computing $\partial_r \Theta$
from the interpolated values would be quite complicated.
\footnote{
	 The arguments of
	 section~\ref{sect-computing-Theta-discrete/geometry-interp}
	 would suggest also having the geometry interpolator
	 guarantee at least $C^0$ continuity of the 2nd~derivative
	 values here, although it's not clear if this would
	 actually be necessary in practice.
	 }

To avoid these problems, I approximate
$\partial_r \Theta$ by a one-sided radial finite difference,
$
\partial_r \Theta
	\approx
	\left[ \Theta(h + \varepsilon) - \Theta(h) \right] \big/ \varepsilon
$,
with $\varepsilon$ typically chosen to be $10^{-6}$
(\cite{Curtis-Reid-1974}, \cite[pp.~266-267]{Stoer-Bulirsch-1980}).
Even though this approximation is only $O(\varepsilon)$ accurate,
in practice this doesn't impair the convergence of Newton's method,
and it's fairly cheap to compute (one extra $\Theta(h)$ evaluation
per Jacobian computation).


\section{Solving the Linear System $\protect\bm{\Jac \cdot \delta h = -\Theta}$}
\label{sect-solving-linear-system}

The Jacobian matrix is an $N_\ang \stimes N_\ang$ sparse matrix;
for typical angular resolutions $N_\ang$ is in the range $300$\,--\,$3000$.
Thus for good efficiency it's important to exploit $\Jac$'s sparsity
in both storage and computation.
I have tried several different linear-equation codes and storage
formats:  For debugging purposes I have found it very useful to store
$\Jac$ as a dense matrix and solve the linear system with \program{LAPACK}
routines.
\footnote{
	 \program{LAPACK} is available from \program{Netlib}
	 ({\tt http://www.netlib.org}).
	 }
$^,$
\footnote{
	 \program{LAPACK}'s condition number estimator is a
	 particularly valuable debugging and diagnostic tool.  For
	 example, incorrect symmetry boundary conditions often
	 result in $\Jac$ being singular (infinite condition number).
	 Another example was in investigating why the Newton iteration
	 sometimes failed to converge in an early version of
	 \program{AHFinderDirect} which used Lagrange rather
	 than Hermite interpolation for the geometry variables
	 (\cf{}~section~\ref{sect-computing-Theta-discrete/geometry-interp}):
	 it was useful to be able to rule out ill-conditioning
	 of the linear system as a possible cause of the
	 convergence failure.
	 }
{}  For better efficiency I now use either an
incomplete--$\sf LU$-decomposition preconditioned conjugate
gradient code \program{ILUCG} (\cite{Kershaw1978:ilucg}),
or the \program{UMFPACK} sparse $\sf LU$-decomposition
code (\cite{Davis-Duff-1997-UMFPACK,Davis-Duff-1999-UMFPACK,
Davis-2002a-UMFPACK-report,Davis-2002b-UMFPACK-report});
\footnote{
	 \program{UMFPACK} is available from
	 {\tt http://www.cise.ufl.edu/research/sparse/umfpack}.
	 }
$^,$
\footnote{
	 \program{UMFPACK} also has a condition number
	 estimator, but as of version~4.0 it appears
	 to be unreliable.
	 }
{} both of these codes use the standard ``compressed row storage''
sparse storage scheme for $\Jac$.  Neither code has been entirely
satisfactory, so I plan to explore other sparse $\sf LU$-decomposition
codes in the near future.


\section{Performance and Accuracy}
\label{sect-performance+accuracy}

In this section I outline the general factors affecting
\program{AHFinderDirect}'s performance (how quickly it can find an AH,
or try to find one) and accuracy (how accurately is an AH found).
I also briefly compare \program{AHFinderDirect} to other AH finders
in these respects.  I defer detailed numerical results to
section~\ref{sect-sample-results}.


\subsection{Performance}
\label{sect-performance+accuracy/performance}

\program{AHFinderDirect}'s performance (the time taken to find, or
try to find, an AH) depends on two main factors:
the total number of angular grid points in the multiple-patch system,
and the number of Newton iterations.  Since there are no computations
done at each Cartesian-grid grid point, the performance is almost
independent of the size and resolution of the Cartesian grid.
\footnote{
	 On an idealized computer there would be no Cartesian-grid
	 resolution dependence at all, but on actual computers cache
	 effects in the geometry interpolator may cause a slight slowdown
	 at higher Cartesian-grid resolutions.
	 }

The total number of angular grid points, $N_\ang$, is determined by
the angular resolution chosen, and whether there are any discrete
symmetries in the multiple-patch system.  Since practical values of
$N_\ang$ vary over roughly an order of magnitude, and empirically the
performance scales very roughly as $N_\ang^{1.4}$, the performance
varies over a wide range from this factor alone.

The number of Newton iterations performed by \program{AHFinderDirect}
is mainly determined by the type of AH being searched for:
\begin{itemize}
\item	\program{AHFinderDirect} is fastest when searching for
	-- and successfully finding -- an AH at each time step of a
	numerical evolution.  In this case the AH typically only
	moves a small distance from one time step to the next,
	so (using the previous time step's position as an
	initial guess for the Newton iteration,
	\cf{}~section~\ref{sect-Newton's-method}) typically only
	3~Newton iterations are needed to locate it at each
	time step.
\item	If \program{AHFinderDirect} finds an AH in an initial
	data slice, typically the initial guess is much less
	accurate, so 6--10~Newton iterations are needed.
\item	\program{AHFinderDirect} is at its slowest when searching
	for -- but failing to find -- an AH at each time step of a
	numerical evolution.  In this case
	(again \cf{}~section~\ref{sect-Newton's-method})
	it typically takes 20~Newton iterations at each time step.
\end{itemize}

As discussed in appendix~\ref{app-multiprocessor-parallelization},
\program{AHFinderDirect} can search for multiple AHs in parallel on
a multiprocessor computer system.  In practice, for large-scale runs
there are usually (many) more processors available than the number of
AHs being searched for.  Assuming this, the elapsed time taken to search
for all the AHs in parallel is basically the maximum of the time taken
to search for each individual AH; this is roughly independent of both
the number of AHs searched for, and the number of processors available.
Part~(b) of the table in appendix~\ref{app-multiprocessor-parallelization}
should make this clearer.


\subsection{Accuracy}

The accuracy with which \program{AHFinderDirect} can find an AH
is mainly determined by the finite differencing errors in the evaluation
of the expansion~$\Theta$.  There are two main error contributions:
the geometry interpolation from the Cartesian grid to the AH position,
and the angular finite differencing within the multiple-patch system
on $S^2$.  (Other error sources such as the interpatch interpolation,
the nonzero $\|\Theta\|$ at which the code considers the Newton iteration
to have ``converged'', and floating-point roundoff errors, are generally
negligible in comparison to the main finite differencing errors.)

For given (smooth) $g_{ij}$ and $K_{ij}$, the errors from the geometry
interpolator are determined by the 3-D (Cartesian) grid spacing $\Delta xyz$,
and by the order of the interpolation scheme.  In the limit of small
$\Delta xyz$, a cubic Hermite geometry interpolator gives $g_{ij}$ and
$K_{ij}$ to $O\bigl((\Delta xyz)^4\bigr)$ and $\partial_k g_{ij}$ to
$O\bigl((\Delta xyz)^3\bigr)$, contributing $O\bigl((\Delta xyz)^3\bigr)$
errors to $\Theta$.
However, at practical resolutions of $\Delta xyz \sim 0.03m$--$0.1m$
I find that the convergence is often $0.5$--$1.0$~power of $\Delta xyz$
better than this, only dropping to the theoretical limits for very
high-resolution grids (in practice, $\Delta xyz \ltsim 0.01m$).

\program{AHFinderDirect} uses 4th~order angular finite differencing
within the multiple-patch system on $S^2$, which contributes
$O\bigl((\Delta \rho\sigma)^4\bigr)$ errors to $\Theta$,
where $\Delta \rho\sigma$ is the angular resolution.


\subsection{Comparison to Other AH-Finding Methods}
\label{sect-performance+accuracy/comparison-to-other-methods}

Curvature-flow or fast-flow methods are widely used for AH-finding
(see, for example,
\cite{Tod91,Kemball91a,Gundlach97a,Shoemaker-Huq-Matzner-2000}).
Conceptually, a flow method starts with a large 2-surface, and
flows this inwards, in such a manner than the flow velocity vanishes
on the AH.  Unfortunately, this means that the method must move
the 2-surface through a large part of the 3-D grid -- and thus must
do nontrivial computations at a large number of 3-D grid points --
before the surface can closely approximate the AH.  In contrast,
an elliptic-equation method such as that used by \program{AHFinderDirect}
need only do computations on a 2-D set of (AH-surface) grid points,
so it can potentially be must faster.

However, a flow method can (at least modulo numerical errors) guarantee
to find the {\em outermost\/} MTS in a slice, whereas an elliptic-equation
method is only locally convergent, and hence offers no information on
what other MTSs might be outside any ``AH'' it finds.

Another common class of AH-finding methods are function-minimization
methods such as those described by \cite{Baumgarte96,Anninos98b}.
These parameterize a trial AH surface by spherical harmonic or other
spectral coefficients, define a surface-integral error norm 
$\int \Theta^2 \, dA$ which has a global minimum of~$0$ at the
AH surface, then use a general-purpose function-minimization
algorithm to minimize the error norm over the surface-coefficient space.
These methods are inherently quite slow because (for a generic slice with
with no continuous symmetries) they must determine a fairly large number
of surface coefficients, and the generic function-minimization algorithm
only ``learns'' a single number (the error norm) for each surface evaluation,
and thus requires many surface evaluations to converge.  For example,
using a spherical harmonic expansion up to order~$N$ to parameterize
the AH surface, there are $O(N^2)$ surface coefficients, so $O(N^2)$
iterations are needed to converge.  Each iteration takes $O(N^2)$ work
to evaluate the surface integral, so the total work is $O(N^4)$.
The exponential convergence of spectral series allows~$N$ to be chosen
to be fairly small for a given surface accuracy, but in practice
function-minimization AH finders are still very slow.

Minimization methods are also inherently somewhat limited in their
accuracy, because the location of the error norm's minimum is very
sensitive to small numerical errors.  (In general, relative errors of
$O(\varepsilon)$ in a smooth function result in relative errors
of $O(\sqrt{\varepsilon})$ in the location of the function's minima.)


\subsection{What makes \program{AHFinderDirect} Fast?}

Based on the above analyses, I think the key algorithm component
which makes \program{AHFinderDirect} fast is the posing of the
AH~equation~\eqref{eqn-Theta(n-i)=0} as an elliptic PDE on $S^2$ 
for the AH shape function~h.  Given this, I believe that any
efficient implementation would result in an AH finder with roughly
the same performance and accuracy as \program{AHFinderDirect}.

A notable example of this is Schnetter's AH finder
(\cite{Schnetter02a,Schnetter03a}), which poses the AH~equation in
the manner as mine, but uses a rather different finite differencing
scheme and solution method for the finite difference equations.
We have not yet made a detailed comparison of our AH finders,
but it appears they are broadly comparable in performance and
accuracy.

Huq's AH finder (\cite{Huq96,Huq00}) also poses the AH~equation as an
elliptic PDE on $S^2$, but he uses Cartesian-grid finite differencing
to evaluate the surface expansion and Jacobian matrix, rather than the
angular-grid finite differencing which Schnetter and I use.  Because
of this, and because he uses numerical perturbations to compute the
Jacobian matrix (\cf{}~section~\ref{sect-Jacobian}), Huq's AH finder
is roughly an order of magnitude slower than mine.


\section{Sample Results}
\label{sect-sample-results}

In this section I present various sample results to test and demonstrate
\program{AHFinderDirect}'s performance.  For comparison, I also show
some results for another AH~finder implemented in the \program{Cactus}
toolkit, the fast-flow method of \cite{Gundlach97a}.
\footnote{
	 Cactus thorn \program{AHFinder}, slightly modified to
	 allow a spherical harmonic expansion up to degree
	 $\ell_{\max} = 50$ (by default the limit is~19).
	 (As discussed below, in practice \program{AHFinder}
	 is limited to $\ell_{\max} \ltsim 20$.)
	 }
{}  (This was the main \program{Cactus} AH finder prior to 
\program{AHFinderDirect}.)
Although some of the test slices are in fact axisymmetric, I
configured both AH~finders to treat the slices as fully 3-D, with
only the discrete symmetries of reflection across the $x$, $y$, and/or
$z=0$~planes as appropriate.  All timings are user-mode CPU times on
a 1.7~GHz dual Pentium~IV processor system (256~KB cache per processor)
with 1.0~GB of memory.


\subsection{Boosted Kerr Slices}
\label{sect-sample-results/boosted-Kerr}

As a first test case, I first consider Kerr spacetime in Kerr-Schild
coordinates (\cite[exercise~33.8]{Misner73}), where the AH is a
coordinate ellipsoid with radia (semi-major axes)
\begin{subequations}
					\label{eqn-Kerr/Kerr-Schild-exact-h}
\begin{eqnarray}
r_z		& = &	(1 + \sqrt{1 - a^2}) m				\\
r_x = r_y	& = &	r_z \sqrt{1 + \left(\dfrac{am}{r_z}\right)^2}
		  =  	\sqrt{\dfrac{2 r_z}{m}} \, m			
\end{eqnarray}
and area
\begin{equation}
A	=	4 \pi (r_z^2 + a^2 m^2)
					\label{eqn-Kerr/Kerr-Schild-exact-A}
\end{equation}
\end{subequations}
where $a = J/m^2$ is the black hole's dimensionless angular momentum.
I then Lorentz-boost this with a velocity $v$~in the $x$~direction.
The horizon area is invariant under the boost, but in the code's
coordinate system, length-contraction makes the AH a triaxial ellipsoid,
and the interaction of the black hole's spin and the boost results
in the slice {\em not\/} being symmetric across either the $x=0$ or
the $y=0$ plane.

Table~\ref{tab-boosted-Kerr-results} shows the accuracy and performance
of \program{AHFinderDirect} and the fast flow AH finder on various
boosted-Kerr slices, for a number of choices of the various numerical
parameters.

\begingroup
\squeezetable
\setlength{\doublerulesep}{0pt}
\begin{table}[tbp]
\begin{center}
\begin{tabular}{D{.}{.}{2}dcccc|rcrdcc|ccrc}
\multicolumn{1}{c}{\P{$a$}}
		& \multicolumn{1}{c}{\P{$v_x$}}
				& \P{$(r_x/\gamma,r_y,r_z)$}
	& \P{$\Delta xyz$}	& \P{origin}			& \P{notes}
	& \multicolumn{6}{c|}{AHFinderDirect}
	& \multicolumn{4}{c}{Fast Flow}					\\
\multicolumn{1}{c}{$a$}
		& \multicolumn{1}{c}{$v_x$}
				& $(r_x/\gamma,r_y,r_z)$
	& $\Delta xyz$		& origin			& notes
	& $\Delta \rho\sigma$	& interp
					& \multicolumn{1}{c}{$N_\ang$}
					      & \multicolumn{1}{c}{\text{time}}
	& $\| \delta h \|_\text{rms}$
				& $(\delta A)/A$
	& $\ell_{\max}$
		& interp
			& time		& $(\delta A)/A$		\\
\hline 
\hline 
0.8		& 0.8		& $(1.07,1.79,1.60)$
	& 0.20			& $(\P{+}0.0,\P{+}0.0)$		&
	& $5^\circ{}\P{.0}$	& H3	& 1121	& 2.0
	& failed		& failed
	& 10	& L3	& \P{0}25	& $1.2 \stimes 10^{-2}$		\\
0.8		& 0.8		& $(1.07,1.79,1.60)$
	& 0.15			& $(\P{+}0.0,\P{+}0.0)$		&
	& $5^\circ{}\P{.0}$	& H3	& 1121	& 4.0
	& $3.4 \stimes 10^{-4}$	& $4.7 \stimes 10^{-4}$
	& 10	& L3	& \P{0}26	& $6.5 \stimes 10^{-3}$		\\
0.8		& 0.8		& $(1.07,1.79,1.60)$
	& 0.10			& $(\P{+}0.0,\P{+}0.0)$		&
	& $5^\circ{}\P{.0}$	& H3	& 1121	& 4.1
	& $5.4 \stimes 10^{-5}$	& $7.8 \stimes 10^{-5}$
	& 10	& L3	& \P{0}33	& $2.2 \stimes 10^{-3}$		\\
0.8		& 0.8		& $(1.07,1.79,1.60)$
	& 0.05			& $(\P{+}0.0,\P{+}0.0)$		&
	& $5^\circ{}\P{.0}$	& H3	& 1121	& 4.2
	& $2.5 \stimes 10^{-5}$	& $4.1 \stimes 10^{-5}$
	& 10	& L3	& \P{0}96	& $4.6 \stimes 10^{-4}$		\\
0.8		& 0.8		& $(1.07,1.79,1.60)$
	& 0.03			& $(\P{+}0.0,\P{+}0.0)$		&
	& $5^\circ{}\P{.0}$	& H3	& 1121	& 4.3
	& $2.6 \stimes 10^{-5}$	& $4.1 \stimes 10^{-5}$
	& 10	& L3	& 350		& $1.0 \stimes 10^{-3}$		\\
\hline 
0.8		& 0.8		& $(1.07,1.79,1.60)$
	& 0.20			& $(\P{+}0.0,\P{+}0.0)$		&
	& $5^\circ{}\P{.0}$	& H2	& 1121	& 1.8
	& $3.6 \stimes 10^{-3}$	& $4.7 \stimes 10^{-3}$
	& 10	& L2	& \P{0}24	& $1.1 \stimes 10^{-2}$		\\
0.8		& 0.8		& $(1.07,1.79,1.60)$
	& 0.15			& $(\P{+}0.0,\P{+}0.0)$		&
	& $5^\circ{}\P{.0}$	& H2	& 1121	& 1.8
	& $4.0 \stimes 10^{-4}$	& $4.3 \stimes 10^{-4}$
	& 10	& L2	& \P{0}26	& $6.2 \stimes 10^{-3}$		\\
0.8		& 0.8		& $(1.07,1.79,1.60)$
	& 0.10			& $(\P{+}0.0,\P{+}0.0)$		&
	& $5^\circ{}\P{.0}$	& H2	& 1121	& 1.9
	& $6.5 \stimes 10^{-4}$	& $1.0 \stimes 10^{-3}$
	& 10	& L2	& \P{0}32	& $2.1 \stimes 10^{-3}$		\\
0.8		& 0.8		& $(1.07,1.79,1.60)$
	& 0.05			& $(\P{+}0.0,\P{+}0.0)$		&
	& $5^\circ{}\P{.0}$	& H2	& 1121	& 1.7
	& $1.3 \stimes 10^{-4}$	& $2.4 \stimes 10^{-4}$
	& 10	& L2	& \P{0}95	& $4.5 \stimes 10^{-4}$		\\
0.8		& 0.8		& $(1.07,1.79,1.60)$
	& 0.03			& $(\P{+}0.0,\P{+}0.0)$		&
	& $5^\circ{}\P{.0}$	& H2	& 1121	& 2.0
	& $3.4 \stimes 10^{-5}$	& $4.7 \stimes 10^{-5}$
	& 10	& L2	& 350		& $1.0 \stimes 10^{-3}$		\\
\hline 
0.8		& 0.8		& $(1.07,1.79,1.60)$
	& 0.05			& $(\P{+}0.0,\P{+}0.0)$		&
	& $7.5^\circ$		& H3	& 533	& 2.0
	& $1.3 \stimes 10^{-4}$	& $2.0 \stimes 10^{-4}$
	& \P{0}7& L2	& 69		& $2.7 \stimes 10^{-3}$		\\
0.8		& 0.8		& $(1.07,1.79,1.60)$
	& 0.05			& $(\P{+}0.0,\P{+}0.0)$		&
	& $5.0^\circ{}$		& H3	& 1121	& 4.2
	& $2.5 \stimes 10^{-5}$	& $4.1 \stimes 10^{-5}$
	& 10	& L2	& \P{0}95	& $4.5 \stimes 10^{-4}$		\\
0.8		& 0.8		& $(1.07,1.79,1.60)$
	& 0.05			& $(\P{+}0.0,\P{+}0.0)$		&
	& $3.0^\circ$		& H3	& 2945	& 13
	& $4.4 \stimes 10^{-6}$	& $7.3 \stimes 10^{-6}$
	& 15	& L2	& 170		& $6.9 \stimes 10^{-4}$		\\
0.8		& 0.8		& $(1.07,1.79,1.60)$
	& 0.05			& $(\P{+}0.0,\P{+}0.0)$		&
	& $1.8^\circ$		& H3	& 7905	& 43
	& $1.3 \stimes 10^{-6}$	& $1.8 \stimes 10^{-6}$
	& 20	& L2	& 280		& $1.3 \stimes 10^{-3}$		\\
0.8		& 0.8		& $(1.07,1.79,1.60)$
	& 0.05			& $(\P{+}0.0,\P{+}0.0)$		&
	& $1.0^\circ$		& H3	& 25025	& 220
	& $9.5 \stimes 10^{-7}$	& $1.3 \stimes 10^{-6}$
	& 28
		& L2	& 2600		& failed			\\
\hline 
0.8		& 0.8		& $(1.07,1.79,1.60)$
	& 0.05			& $(-0.5,-0.9)$			& (a)
	& $5^\circ{}\P{.0}$	& H3	& 1121	& 5.7
	& $2.4 \stimes 10^{-5}$	& $3.3 \stimes 10^{-5}$
	& 10	& L2	& 960	& $1.2 \stimes 10^{-3}$			\\
0.8		& 0.8		& $(1.07,1.79,1.60)$
	& 0.05			& $(\P{+}0.0,\P{+}0.0)$		&
	& $5^\circ{}\P{.0}$	& H3	& 1121	& 4.2
	& $2.5 \stimes 10^{-5}$	& $4.1 \stimes 10^{-5}$
	& 10	& L2	& \P{0}95	& $4.5 \stimes 10^{-4}$		\\
0.8		& 0.8		& $(1.07,1.79,1.60)$
	& 0.05			& $(+0.5,+0.9)$			& (a)
	& $5^\circ{}\P{.0}$	& H3	& 1121	& 7.2
	& $1.7 \stimes 10^{-5}$	& $2.7 \stimes 10^{-6}$
	& 10	& L2	& 960	& $5.6 \stimes 10^{-3}$			\\
\hline 
\hline 
0.99		& 0.8		& $(0.91,1.51,1.14)$
	& 0.03			& $(\P{+}0.0,\P{+}0.0)$		&
	& $3^\circ{}\P{.0}$	& H3	& 2945	& 15
	& $3.3 \stimes 10^{-6}$	& $4.8 \stimes 10^{-6}$
	& 15	& L2	& 2300		& failed			\\
0.999		& 0.8		& $(0.87,1.45,1.04)$
	& 0.03			& $(\P{+}0.0,\P{+}0.0)$		&
	& $3^\circ{}\P{.0}$	& H3	& 2945	& 16
	& $9.5 \stimes 10^{-6}$	& $1.4 \stimes 10^{-5}$
	& 15	& L2	& 1600		& failed			\\
0.999		& 0.95		& $(0.45,1.45,1.04)$
	& 0.02			& $(\P{+}0.0,\P{+}0.0)$		& (b,d)
	& $3^\circ{}\P{.0}$	& H3	& 2945	& 13
	& $4.0 \stimes 10^{-4}$	& $6.5 \stimes 10^{-4}$
	& \multicolumn{4}{c}{not tested on this slice}			\\
0.999		& 0.98		& $(0.29,1.45,1.04)$
	& 0.02			& $(\P{+}0.0,\P{+}0.0)$		& (c,d)
	& $3^\circ{}\P{.0}$	& H3	& 2945	& 12
	& $1.3 \stimes 10^{-4}$	& $1.1 \stimes 10^{-4}$
	& \multicolumn{4}{c}{not tested on this slice}			\\
\hline 
\hline 
\end{tabular}
\end{center}
\footnotesize
\footnotetext[1]{
		Fast-flow initial guess changed to
		a coordinate sphere of radius~$2.5$,
		and Cartesian grid enlarged to size $\pm 4$
		(the larger Cartesian grid size points should have only
		minimal effects on \program{AHFinderDirect}'s performance,
		but should slow the fast flow method by a factor
		of~$(4/2.5)^3 \approx 4$)
		}
\footnotetext[2]{
		\program{AHFinderDirect} initial guess changed to
		a coordinate ellipsoid of radia $(0.5, 1.5, 1.0)$
		}
\footnotetext[3]{
		\program{AHFinderDirect} initial guess changed to
		a coordinate ellipsoid of radia $(0.3, 1.5, 1.0)$
		}
\footnotetext[4]{
		Cartesian grid shrunk to size~$\pm 2$
		to reduce memory usage (this should have only
		minimal effects on \program{AHFinderDirect}'s performance)
		}
\begin{flushleft}
Table columns not described in the main text:\\[1ex]
\begin{tabular}{ll}
origin		& The $(x,y)$ components of the local coordinate origin
		  (the $z$~component is always~0)			\\
interp		& The geometry interpolator:				\\
		& H(L) means Hermite (Lagrange)
		  polynomial interpolation,				\\
		& the following integer gives the order			\\[1ex]
time		& user-mode CPU time in seconds				\\[1ex]
$\| \delta h \|_\text{rms}$
		& The rms-norm over the angular grid
		  of the error in the computed AH radius $h$		\\[1ex]
$(\delta A)/A$	& The relative error in the computed AH area		\\[1ex]
$\ell_{\max}$	& The maximum order of the spherical harmonic expansion
\end{tabular}
\end{flushleft}
\caption[Boosted-Kerr Test Results]
	{
	This table shows the accuracy and performance of
	\program{AHFinderDirect} on various boosted Kerr slices.
	In each case the black hole has dimensionless rest mass $m = 1$.
	Except as noted, the Cartesian grid is of size $\pm 2.5$
	(more precisely, $[-2.5,+2.5]$ in $x$ and $y$ and $[0,2.5]$ in $z$,
	with $z \leftrightarrow -z$ reflection symmetry across the $z=0$ plane).
	Except as noted, the \program{AHFinderDirect} initial guess
	is a coordinate sphere of radius~$1.5$,
	and the fast-flow initial guess is
	a coordinate sphere of radius~$2$.
	\program{AHFinderDirect} used the \program{ILUCG} sparse matrix
	routines in all cases.
	In most cases the $\infty$-norm error in the \program{AHFinderDirect}
	AH shape was less than twice the rms-norm error shown here;
	in no case did it exceed 5~times the rms-norm error.
	}
\label{tab-boosted-Kerr-results}
\end{table}
\endgroup

The first section of the table shows \program{AHFinderDirect}'s
behavior as the resolution of the underlying Cartesian grid is varied,
using the default cubic Hermite geometry interpolator.  At very low
resolution ($\Delta xyz = 0.2$) \program{AHFinderDirect} fails to
find the AH, due to the geometry interpolation ``seeing'' the Kerr
ring singularity.  At higher resolution (decreasing $\Delta xyz$)
the accuracy improves rapidly, until it levels out at high resolutions
due to the angular finite differencing errors.  For the computer system
used here, the time taken to find the AH is essentially independent
of the Cartesian grid resolution.

The second section of the table shows \program{AHFinderDirect}'s
behavior as the resolution of the underlying Cartesian grid is varied,
using a lower-order (quadratic) geometry interpolator.
Compared to the default (cubic) geometry interpolator,
this makes \program{AHFinderDirect} a factor of~2 to~3 faster,
and roughly an order of magnitude less accurate.
Also, at the very lowest resolution \program{AHFinderDirect} is
now able to find the AH, when it couldn't find it using the cubic
interpolator.

Comparing the first two sections of the table shows that changing
the interpolation order seems to make only a minor difference to the
fast flow method's behavior; all the remaining tests use its default
(quadratic Lagrange) geometry interpolator.  As discussed in
section~\ref{sect-performance+accuracy/comparison-to-other-methods},
the fast flow method becomes much slower at high Cartesian grid
resolutions.

The third section of the table shows \program{AHFinderDirect}'s
behavior as the angular resolution is varied.  As the resolution is
increased (decreasing $\Delta \rho\sigma$) \program{AHFinderDirect}
becomes slower but more accurate, until the error levels off at
high angular resolutions due to the Cartesian-grid geometry
interpolation errors.

The third section of the table also shows the fast flow method
becoming slower as its resolution parameter $\ell_{\max}$ is increased.
Unfortunately, beyond $\ell_{\max} \approx 10$ the method's accuracy
stops improving and begins to worsen, and beyond $\ell_{\max} \approx 20$
the fast flow method fails to find the AH.  I suspect this is due to
numerical ill-conditioning, but I haven't investigated this in detail.

The fourth section of the table shows \program{AHFinderDirect}'s
behavior when the local coordinate origin is offset from the coordinate
origin.  Notice that the accuracy with which the AH is found isn't
significantly changed, and the time taken to find the AH is only
mildly increased, even when the local coordinate origin is offset
by up to $\thalf$ the AH radius.  The fast flow method is still
able to find the AH with the offset local coordinate origins, but
it requires changes to the initial guess, and (even after correcting
for the larger grid) it slows dramatically and becomes less accurate.

The final section of the table shows \program{AHFinderDirect}'s
behavior on some more difficult boosted-Kerr slices, where the spin
is closer to maximal and/or the boost is larger.  Because the ring
singularity in Kerr moves closer to the AH at high spins, and
length contraction makes the AH strongly triaxial at high boosts,
these tests used higher Cartesian and angular resolutions than
the previous tests.  \program{AHFinderDirect} still finds the
horizon rapidly and with high accuracy in these cases, although
in the two most difficult cases quite good initial guesses were
required.  The fast flow method is not able to find the AH for
any of these cases, even with some adjustment of its initial guesses
(this may be due in part to its user interface only allowing for
axisymmetric initial guesses).

Across all the boosted-Kerr tests, \program{AHFinderDirect} is
roughly an order of magnitude faster, and two orders of magnitude
more accurate, than the fast flow method.


\subsection{Misner and Brill-Lindquist Slices}
\label{sect-sample-results/Misner+Brill-Lindquist}

The Misner (\cite{Misner60,Misner63}) and Brill-Lindquist (\cite{Brill63})
initial data slices are standard test problems in numerical relativity.
Both are time-symmetric ($K_{ij} = 0$),
3-conformally-flat ($g_{ij} = \Psi \eta_{ij}$ for some
spatially-varying conformal factor~$\Psi$), and
(for suitable values of their parameters)
may contain any number $N \ge 1$ of black holes.

The simplest case of Misner data (and the only case I consider here)
is that of 2~throats, each of bare mass unity.  Here the conformal factor
is
\begin{equation}
\Psi = 1 + \sum_{n=1}^\infty
	   \frac{1}{\sinh (n \mu)}
	   \left( \frac{1}{r_n^+} + \frac{1}{r_n^-} \right)
					\label{eqn-Misner-conformal-factor}
\end{equation}
where
\begin{equation}
r_n^\pm	= \sqrt{x^2 + y^2 + \left[ z \pm \coth (n\mu) \right]^2}
					\quad \text{.}	
\end{equation}
with $\mu > 0$ is a real parameter.  The individual throats are located
at coordinate positions $(0, 0, \pm \coth \mu)$.  For small~$\mu$ there
is only a single AH enclosing both throats, while for large~$\mu$ there
are individual AHs enclosing each throat, but no common AH enclosing both
throats.

The conformal factor for $N$-throat Brill-Lindquist initial data is
\begin{equation}
\Psi(\vect{x})
	= 1 + \half \sum_{i=1}^N
	      \frac{m_i}{\left| \vect{x} - \vect{x_i} \right|}
\end{equation}
where the $i$th throat has bare mass $m_i$ and is located at the
coordinate position $\vect{x_i}$.  Here I consider the cases $N=2$
and $N=3$, where the throats each have bare mass unity, and are
uniformly spaced in a coordinate circle of radius~$R > 0$.  Similarly
to the Misner data, for small~$R$ there is a single common AH, while
for large~$R$ there are $N$~individual AHs but no common AH.

The AH-finder test problem I consider here is to
numerically determine the \defn{critical} value of the parameter
($\mu$~for Misner, $R$~for Brill-Lindquist)
at which the common horizon appears/disappears for each family of slices.
To do this, I used the \program{Cactus} thorn \program{IDAnalyticBH}
to construct the initial data slices, approximating the infinite
sum~\eqref{eqn-Misner-conformal-factor} by its first 30~terms.
\footnote{
	 Raising this to 50~terms changed the numerically
	 computed critical $\mu$ by $< 10^{-12}$,
	 and the horizon area by $< 10^{-7}$.
	 }
{}  For each of a number of combinations of the \program{Cactus}
Cartesian grid spacing and the \program{AHFinderDirect} angular
grid spacing,
\footnote{
	 I used very high resolutions here, with grid
	 spacings as small as $0.01$ for the \program{Cactus}
	 3-D grid and $0.5^\circ$ for the
	 \program{AHFinderDirect} angular grid.
	 }
{} I used a continuation-method binary search (described in detail
in appendix~\ref{app-continuation-binary-search}) to determine the
critical parameter.  I did convergence tests (\cite{Choptuik91})
in both grid spacings to verify that the values shown are reliable
estimates of the true continuum values, and I used Richardson
extrapolation in the angular grid spacing to improve the accuracy.
\footnote{
	 Because the \program{AHFinderDirect} angular grid
	 isn't commensurate with the \program{Cactus} Cartesian
	 grid, the geometry-interpolation errors effectively
	 have a quasirandom phase at each angular grid point.
	 This prevents these errors from being smooth enough
	 to allow Richardson extrapolation on the Cartesian
	 grid spacing.  However, the variation of the computed
	 critical parameters with the Cartesian grid spacing
	 can still be used qualitatively to help estimate the
	 critical parameters' accuracy.
	 }
{}  Table~\ref{tab-Misner+BL-critical-mu} shows the results,
together with values reported by \cite{Alcubierre98b} for comparison.
The \program{AHFinderDirect} values are in excellent agreement with
those of \cite{Alcubierre98b}, and are dramatically more accurate.

\begin{table}[tbp]
\def\Aname{Alcubierre {\it et al.}}
\def\Cname{\v{C}ade\v{z}}
\setlength{\tabcolsep}{0.5em}
\begin{center}
\begin{tabular}{lcccc}
				&
	& \Aname
	& \multicolumn{2}{c}{\program{AHFinderDirect}}			\\
Test Problem			& parameter
	& critical parameter
	& critical parameter			& critical AH area	\\
\hline 
Misner				& $\mu$
	& $1.364\P{\spm}$
	& $1.365\,071\,172 \spm 3$		& $409.549\,358 \spm 3$	\\
Brill-Lindquist	2-throat	& $R$
	& $0.766\P{\spm}$
	& $0.766\,197\,45 \spm 5\P{0}$	& $196.407\,951 \spm 3$	\\
Brill-Lindquist	3-throat	& $R$
	& $1.18 \spm 4$
	& $1.195\,499\,53 \spm 5\P{0}$	& $444.756\,224 \spm 3$
\end{tabular}
\end{center}
\caption[Critical Parameters for 2- and 3-throat
	 Misner and Brill-Lindquist Slices]
	{
	This table shows the maximum Misner~$\mu$ and
	Brill-Lindquist~$R$ for which \program{AHFinderDirect}
	found a common AH, along with the area of that common AH.
	All uncertainties are in units of the last digits shown.
	For the 2-throat Brill-Lindquist data, other values
	in the literature include
	$R = 0.767 \spm 1$ (\protect\cite{Cadez74})
	and
	$R = 0.768$ (\protect\cite{Shoemaker-Huq-Matzner-2000}).
	However, \protect\cite{Shoemaker-Huq-Matzner-2000}
	report a critical AH area for this case of
	$184.16$, about $6\%$~different from \program{AHFinderDirect}'s.
	I don't know the cause of this discrepancy.
	}
\label{tab-Misner+BL-critical-mu}
\end{table}

\subsection{Binary Black Hole Collision Spacetimes}
\label{sect-sample-results/binary-BH}

\begin{figure}[tbp]
\begin{center}
\includegraphics{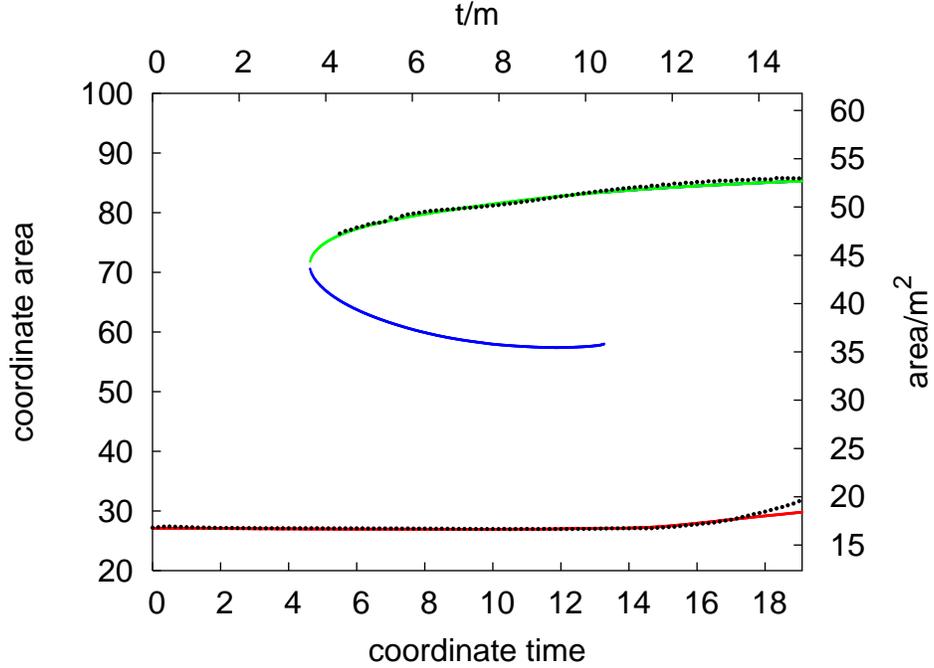}
\end{center}
\caption[Apparent Horizon Areas in Misner $\mu = 2.0$ Collision]
	{
	This figure shows the areas of the various AHs
	in the Misner $\mu = 2.0$ collision described
	in table~\protect\ref{tab-misner12-066-parameters}.
	The black points are the areas found by the fast flow AH finder;
	the other curves are all from \program{AHFinderDirect}.
	The gradual rise in the area of the outer common AH
	after $t \approx 9$, and in the area of the individual
	AH after $t \approx 15$, is due to outer boundary reflections
	making the overall evolution inaccurate.
	}
\label{fig-misner12-066/AH-areas}
\end{figure}

\begin{figure}[tbp]
\begin{center}
\begin{picture}(160,100)
\put(6,10){
	  \begin{picture}(0,0)
	  \put(15,90){Part~(a)}
	  \put(-23.65,-44.10){\includegraphics[scale=1.25]{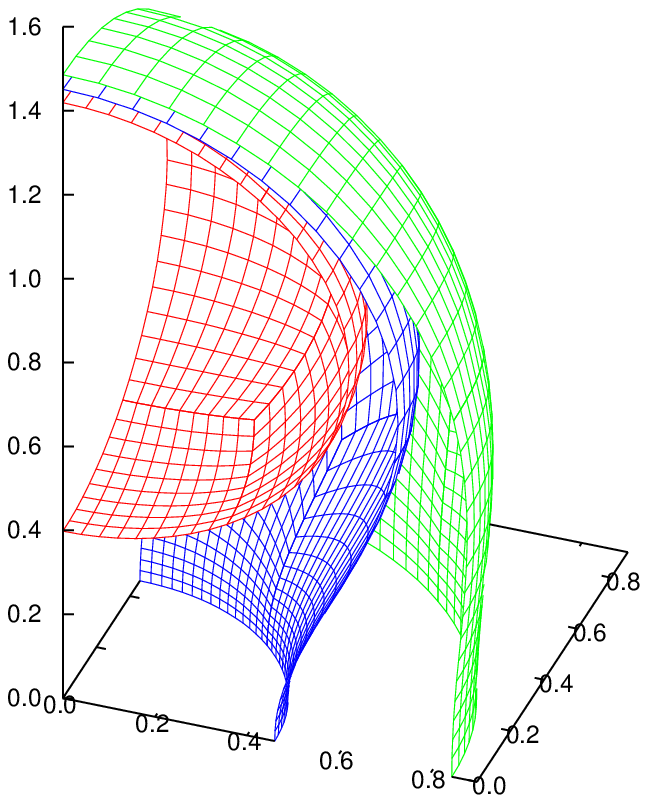}}
	  \end{picture}
	  }
\put(88,10){
	   \begin{picture}(0,0)
	   \put(15,90){Part~(b)}
	   \put(-23.65,-44.10){\includegraphics[scale=1.25]{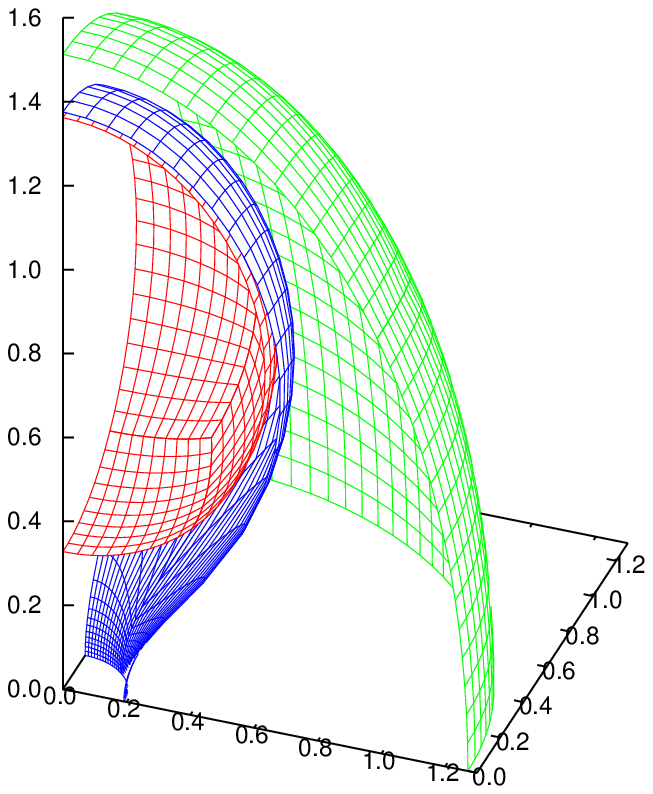}}
	   \end{picture}
	   }
\end{picture}
\end{center}
\caption[Apparent Horizons at $t=5.00$ ($3.93m$) Misner $\mu = 2.0$ Collision]
	{
	This figure shows the three AHs
	in the Misner $\mu = 2.0$ collision described
	in table~\protect\ref{tab-misner12-066-parameters}.
	Part~(a) shows the horizons at $t = 5.00$ ($3.93m$);
	part~(b) shows them at $t = 8.00$ ($6.28m$).
	In both parts the color coding matches that of
	figure~\protect\ref{fig-misner12-066/AH-areas}.
	}
\label{fig-misner12-066/AHs-at-t=5-and-8}
\end{figure}

As a final example, I consider the binary black hole collision evolution
described in table~\ref{tab-misner12-066-parameters}.
Figure~\ref{fig-misner12-066/AH-areas} shows the AH areas found by
\program{AHFinderDirect} and the fast flow AH finder for this evolution.
For the AHs they both find, the two AH finders agree very well.
\program{AHFinderDirect} found the outer common AH
somewhat sooner than the fast flow AH finder
($t=4.633$ ($3.64m$) versus $t=5.50$ ($4.32m$)),
and was the only finder to find the inner common AH.
Figure~\ref{fig-misner12-066/AHs-at-t=5-and-8} shows the 3~AHs found by
\program{AHFinderDirect} at two times during the evolution.

For this evolution the mean CPU times per time step were
$5.2$ seconds for \program{AHFinderDirect} and (for those time
steps for which it ran) $55$~seconds for the fast flow AH finder,
so despite searching for 3~AHs instead of~2,
\program{AHFinderDirect} was about an order of magnitude faster
than the fast flow method.

Since the runs just described, I have changed \program{AHFinderDirect}'s
default geometry interpolator from cubic to quadratic Hermite.  In
practice \program{AHFinderDirect} is usually used in numerically
computed slices whose geometries have numerical errors large enough
to dominate \program{AHFinderDirect}'s intrinsic errors.  Thus
(\cf{}~section~\protect\ref{sect-sample-results/boosted-Kerr}) the
lower-order geometry interpolation makes little difference to the
practical accuracy with which the apparent horizons are found, and
it speeds up \program{AHFinderDirect} by roughly a factor of~2 to~3.
For example, in a recent large binary-black-hole collision simulation
(details of which will be reported elsewhere), \program{AHFinderDirect}
(using the \program{UMFPACK} sparse matrix routines) averaged
$1.7$~seconds per time step, as compared with $61$~seconds per
time step for the fast flow AH finder.

\begin{table}[tbp]
\begin{center}
\setlength{\tabcolsep}{0.5em}
\begin{tabular}{ll}
\hline 
Initial data	& Misner $\mu = 2.0$ (ADM mass $m = 1.272$)		\\
		& black holes located on the $z$~axis
		  at $z = \pm 0.99$					\\
Coordinates	& $1 \splus \log$ lapse, $\Gamma$-driver shift		\\
Numerical grid	& $\Delta xyz = 0.0666$ ($0.052m$);
		  $+xyz$~octant symmetry				\\
Time integration& iterated Crank-Nicholson (3 iterations)		\\
		& Courant number $\Delta t / \Delta xyz = 0.25$		\\
Outer boundaries& in the computational coordinates the outer boundaries	\\
		& were at $xy_{\max} = 5.37$ ($4.22m$),
		  $z_{\max} = 6.43$ ($5.05m$);				\\
		& a ``fisheye'' nonuniform-grid transformation was used	\\
		& which placed the physical outer boundaries at		\\
		& $xy_{\max} = 13.1$ ($10.3m$),
		  $z_{\max} = 16.3$ ($12.8m$)				\\
\program{AHFinderDirect}
		& searching for individual and inner/outer common AHs
		  at each time step					\\
		& cubic Hermite geometry interpolation			\\
		& angular resolution $\Delta \rho\sigma = 5^\circ$
		  for individual AH					\\
		& \quad
		  ($N_\ang = 580$, $+xy$~quadrant symmetry,
		   local coordinate origin $(0,0,0.85)$)		\\
		& angular resolution $\Delta \rho\sigma = 3^\circ$
		  for inner and outer common AHs			\\
		& \quad
		  ($N_\ang = 768$, $+xyz$~octant symmetry,
		   local coordinate origin $(0,0,0)$)			\\
		& \program{UMFPACK} sparse matrix routines		\\
Fast flow AH finder
		& searching for individual and common AHs
		  each 10 time steps					\\
		& fast flow method (\cite{Gundlach97a}),
		  spherical harmonics up to $\ell_{\max} = 10$		\\
		& same local coordinate origins
		  as \program{AHFinderDirect}				\\
\hline 
\end{tabular}
\end{center}
\caption[Run Parameters for Misner Collision Evolution]
	{
	This table gives various parameters for the binary
	black hole collision evolution shown in
	figures~\protect\ref{fig-misner12-066/AH-areas}
	and~\protect\ref{fig-misner12-066/AHs-at-t=5-and-8}.
	}
\label{tab-misner12-066-parameters}
\end{table}

\section{Conclusions}

In this paper I present a detailed description of a new numerical
apparent horizon finder for 3-dimensional Cartesian grids,
\program{AHFinderDirect}.
\program{AHFinderDirect} is typically at least an order of magnitude
faster than other widely-used apparent horizon finders; in particular
\program{AHFinderDirect} is fast enough that it's practical to find
apparent horizons at each time step of a numerical evolution.
This allows apparent horizon positions to readily be used in
coordinate conditions
(see, for example,
\cite{Sperhake-etal-2003-densitized-lapse-wobbling-BH-evolution})
or for other diagnostic purposes.

\program{AHFinderDirect} is also very accurate, typically finding
apparent horizons to within $\sim 10^{-5} m$ in coordinate position.

\program{AHFinderDirect} is implemented within the \program{Cactus}
computational toolkit, and is freely available (GNU GPL licensed)
by anonymous CVS checkout from \verb|cvs.aei.mpg.de:/numrelcvs|
in the directory \verb|AEIThorns/AHFinderDirect|.
It would also be fairly easy to port \program{AHFinderDirect} to a
different (non-\program{Cactus}) numerical relativity code.


\appendix

\section{Details of the Multiple Patch System}
\label{app-multiple-patch-details}


\subsection{Coordinates}
\label{app-multiple-patch-details/coordinates}

I define angular coordinates on $S^2$ based on rotation angles about
the local $xyz$~coordinate axes:
\begin{equation}
							\label{eqn-mu-nu-phi}
\renewcommand{\arraystretch}{1.25}
\begin{array}{c@{}l@{}l}
\mu	& {} = \text{rotation angle about the local $x$ axis}
	& {} = \arctan(y/z)						\\
\nu	& {} = \text{rotation angle about the local $y$ axis}
	& {} = \arctan(x/z)						\\
\phi	& {} = \text{rotation angle about the local $z$ axis}
	& {} = \arctan(y/x)						
\end{array}
\end{equation}
where all the arctangents are 4-quadrant based on the signs of $x$, $y$,
and $z$.  I then define coordinate patches covering neighborhoods of the
$\pm z$, $\pm x$, and $\pm y$ axes, using the generic patch coordinates\\[-2ex]
\begin{equation}
\renewcommand{\arraystretch}{1.25}
\begin{tabular}{l@{}l}
$\pm z$	& \ patch has generic patch coordinates
	  $(\rho,\sigma) = (\mu,\nu)$					\\
$\pm x$	& \ patch has generic patch coordinates
	  $(\rho,\sigma) = (\nu,\phi)$					\\
$\pm y$	& \ patch has generic patch coordinates
	  $(\rho,\sigma) = (\mu,\phi)$					
\end{tabular}
						\label{eqn-patch-rho-sigma}
\end{equation}

The resulting set of 6~patches cover $S^2$ without coordinate
singularities.
\footnote{
	 Another way to visualize these patches and
	 coordinates is to imagine an $xyz$~cube with
	 $xyz$~grid lines painted on its face.  Now
	 imagine the cube to be flexible, and inflate
	 it like a balloon, so it becomes spherical
	 in shape.  The resulting coordinate lines
	 will closely resemble those for $(\mu,\nu,\phi)$
	 coordinates.
	 }
{}  Alternatively, if the slice has $z \leftrightarrow -z$
reflection symmetry about the local coordinate origin, then the
5~patches $+z$, $\pm x$, and $\pm y$ cover the $+z$~hemisphere of
$S^2$.  Similarly, suitable sets of~4 or 3~patches may be used to
cover quadrants or octants of $S^2$ respectively; figure~\ref{fig-3-patch}
shows an example of this last case.


\subsection{Ghost Zones}
\label{app-multiple-patch-details/ghost-zones}

Each patch is a rectangle in its own $(\rho,\sigma)$ coordinates;
I use the usual \defn{ghost zone} technique for handling finite
differencing near the patch boundaries.  I refer to the non-ghost-zone
part of a patch's grid as its \defn{nominal} grid.  Adjacent patches'
nominal grids just touch.  (Grid-function values in) the ghost zones
are filled in from values in their own and other patches' nominal grids
by symmetry operations and/or interpatch interpolations.

With the coordinate choice~\eqref{eqn-patch-rho-sigma}, adjacent
patches always share the angular coordinate perpendicular to their
mutual boundary, so the interpatch interpolations need only be done
in one dimension, in the direction parallel to the boundary.  Since
off-centering an interpolant, particularly a high-order one,
significantly degrades its accuracy, I have tried to design the
algorithms to keep the interpolations centered wherever possible.
\footnote{
	 Another reason to keep the interpolations
	 centered in \program{AHFinderDirect} was to
	 allow re-use of the multiple-patch software
	 from an earlier time evolution code
	 (\cite{Thornburg2000:multiple-patch-evolution}),
	 where centering the interpolations helped
	 keep the evolution stable.
	 }

\begin{figure}[tbp]
\begin{center}
\begin{picture}(160,160)
\put(25,110){
	    \begin{picture}(0,0)
	    \put(-5,37.5){Part~(a)}
	    \put(-16.75,-16.70){\includegraphics{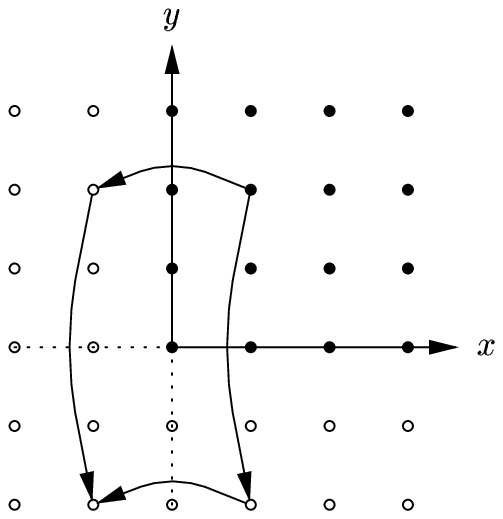}}
	    \end{picture}
	    }
\put(110,110){
	     \begin{picture}(0,0)
	     \put(-5,37.5){Part~(b)}
	     \put(-29.10,-20.10){\includegraphics{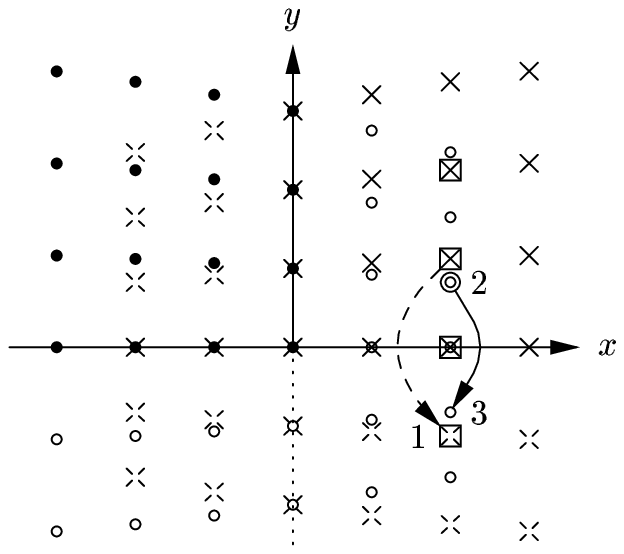}}
	     \end{picture}
	     }
\put(40,30){
	   \begin{picture}(0,0)
	   \put(0,0){\circle{2}}
	   \put(-5,50){Part~(c)}
	   \put(-35.25,-30.95){\includegraphics{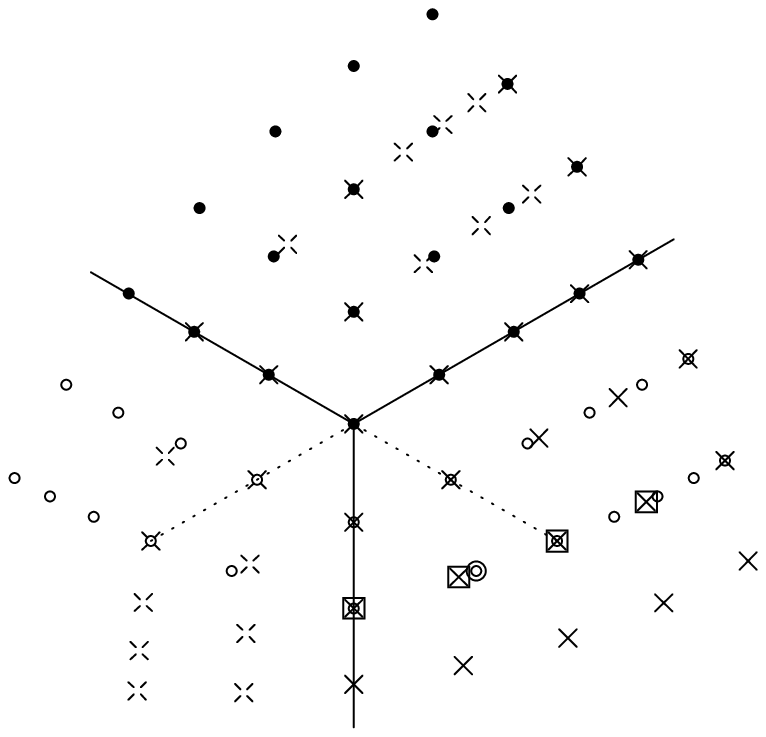}}
	   \end{picture}
	 }
\put(95,0){
	   \begin{picture}(0,0)
	   \put(0,0){\includegraphics{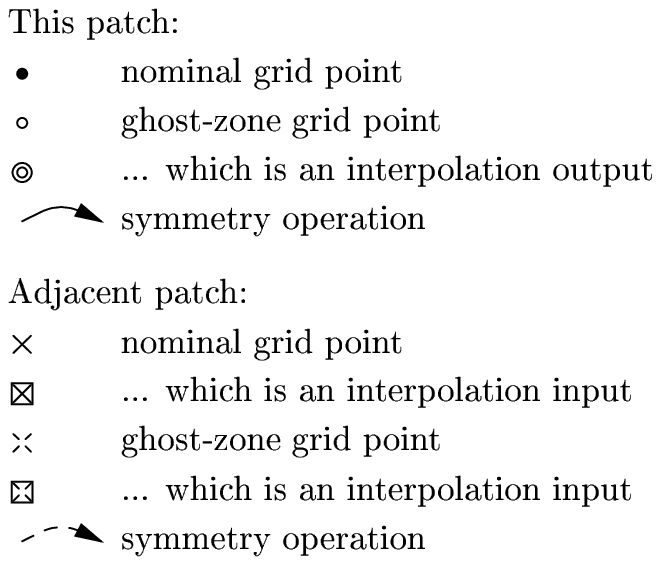}}
	   \end{picture}
	   }
\end{picture}
\end{center}
\caption[Interpatch Interpolation in the Corners of a Ghost Zone]
	{
	This figure shows the combinations of symmetry operations
	and interpatch interpolations used to fill in grid function
	values in ghost-zone corners.
	Part~(a)
	(where there are both $x \leftrightarrow -x$
	and $y \leftrightarrow -y$ reflection symmetries)
	shows a corner between two symmetry ghost zones.
	Part~(b)
	(where there is a $y \leftrightarrow -y$ reflection symmetry)
	shows a corner between a symmetry and an interpatch ghost zone;
	sample output points for each phase of the 3-phase algorithm
	described in the text are labelled as 1, 2, and 3.
	Part~(c) (where there are no symmetries)
	shows a corner between two interpatch ghost zones
	(this only happens when 3~patches meet at a corner).
	In each part, arrows show the symmetry operations,
	and for parts~(b) and~(c) the boxed and circled points
	show the inputs and outputs for the interpatch interpolations.
	}
\label{fig-interpatch-corners}
\end{figure}

The most complicated part of the multiple-patch scheme is in the
handling of the \defn{corner} ghost-zone grid points, those ghost-zone
grid points which are outside their patch's nominal grid in both the
$\rho$ and the $\sigma$ directions.
Figure~\ref{fig-interpatch-corners} shows the three basic cases:
\begin{description}
\item[{\rm (a)}]
	Figure~\ref{fig-interpatch-corners}(a) shows an example of
	a corner between two symmetry ghost zones.
	In this case it takes 2~sequential symmetry operations
	(shown by the curved arrows) to fill in the corner from the
	nominal grid.  Fortunately, symmetry operations commute,
	\ie{} the results are independent of the order of the two
	symmetry operations.
\item[{\rm (b)}]
	Figure~\ref{fig-interpatch-corners}(b) shows an example of
	a corner between a symmetry and an interpatch ghost zone.
	To keep the interpolations centered, I use a 3-phase algorithm here:
	\begin{enumerate}
	\item	Use symmetry operations
		(for example, the one shown by the dotted arrow in the figure)
		to fill in the non-corner ghost-zone points
		in the neighboring patch.
	\item	Do a centered interpatch interpolation from the
		neighboring patch to this patch; this interpolation
		may use some points from the neighboring patch's
		ghost zone.
	\item	Use symmetry operations
		(for example, the one shown by the solid arrow in the figure)
		to fill in the corner ghost-zone points in this patch.
	\end{enumerate}
\item[{\rm (c)}]
	Figure~\ref{fig-interpatch-corners}(c) shows an example of
	a corner between two interpatch ghost zones
	(this only happens when 3~patches meet at a corner).
	This case requires only a single interpatch interpolation
	for each ghost-zone grid point.
\end{description}

\program{AHFinderDirect} actually uses the following 3-phase algorithm
(which includes each of (a)--(c) above as special cases) to perform all
the necessary symmetry operations and interpatch interpolations across
all patches, in a correct order:
\footnote{
	 The ordering of the phases is essential to
	 obtain correct results, within each phase
	 the different ghost zones and patches may
	 be processed in any order.
	 }
\begin{enumerate}
\item	Use symmetry operations to fill in the non-corner
	parts of all symmetry ghost zones in all patches.
\item	Use interpatch interpolations to fill
	in all interpatch ghost zones in all patches.
\item	Use symmetry operations to fill in the corners
	of all symmetry ghost zones in all patches.
\end{enumerate}

\subsection{Jacobian Computation}
\label{app-multiple-patch-details/Jacobian}

The symbolic-differentiation Jacobian~\eqref{eqn-Jacobian} must be
modified to take into account the ghost-zone symmetry operations and
interpatch interpolations described in the previous subsection.  This
is essentially a straightforward application of the chain rule for each
of the $\Dmol_u$ and $\Dmol_{uv}$ terms in~\eqref{eqn-Jacobian}.
Figure~\ref{fig-Jacobian-algorithm} shows the resulting algorithm
in detail.

\begin{figure}[tbp]
\def\assign{\leftarrow}
\def\ttt{\P{$\Jac \assign {}$}}
\begin{tabbing}
$\Jac \assign 0$ matrix							    \\
\ttt\=for each angular grid point $\I$					  \+\\
    \{									    \\
    \ttt\=for each angular coordinate index $u$
	  and pair of indices $uv$					  \+\\
	\{								    \\
	\ttt\=for each molecule index $\m \in \Dmol_u$
	      or $\Dmol_{uv}$ respectively	  \+\\
	    \{								    \\
	    $\J \assign \I+\m$						    \\
	    $\displaystyle
	     \text{temp}
	        \assign
		   \frac{\partial \Theta}{\partial (\partial_u h)}
		   \Dmol_u[\m]$
	    or
	    $\displaystyle
	          \frac{\partial \Theta}{\partial (\partial_{uv} h)}
		  \Dmol_{uv}[\m]$
	    respectively						    \\
	    if \=($\J \in \text{nominal grid of the patch containing $\I$}$)
									  \+\\
	       then \=$\Jac_{\I\J} \assign \Jac_{\I\J} + \text{temp}$	    \\
	       else \>\{						  \+\\
		      \ttt\=for each angular grid point $\K$
			    used in computing $h[\J]$			  \+\\
			    via the 3-phase algorithm of
			appendix~\ref{app-multiple-patch-details/ghost-zones}
									    \\
			    \{						    \\
			    $\displaystyle
			     \Jac_{\I\K} \assign
				\Jac_{\I\K}
				+ \text{temp}
				  \times \left(
					 \frac{\partial h[\J]}{\partial h[\K]}
					 \text{~for the 3-phase algorithm}
					 \right)$			    \\
			    \}						  \-\\
		      \}						\-\-\\
	    \}								  \-\\
	\}								  \-\\
    $\displaystyle
     \Jac_{\I\I} \assign \Jac_{\I\I}
			 + \frac{\Theta(h + \varepsilon) - \Theta(h)}
				{\varepsilon}$				    \\
    \}									
\end{tabbing}
\caption[Symbolic-Differentiation Jacobian Algorithm Including Ghost Zones]
	{
	This figure shows \program{AHFinderDirect}'s
	overall Jacobian-computation algorithm, including
	ghost-zone handling.
	}
\label{fig-Jacobian-algorithm}
\end{figure}


\section{Multiprocessor and Parallelization Issues}
\label{app-multiprocessor-parallelization}

\section{Searching for the Critical Parameter
	 of a 1-Parameter Initial Data Sequence}
\label{app-continuation-binary-search}

In the interests of brevity these two appendices are omitted here.
They appear as supplemental material in the online edition
of this paper, and as appendices~B and~C in the preprint-archive
version (gr-qc/0306056).


\section*{Acknowledgements}

I thank the Alexander von Humboldt foundation,
the AEI visitors program,
and the AEI postdoctoral fellowship program
for financial support.
I thank Peter Diener and Ian Hawke for useful conversations.
I think Peter Diener, Ian Hawke, Scott Hawley, Denis Pollney,
Ed Seidel, and Erik Schnetter for helpful comments on various
drafts of this paper.
I thank two anonymous referees for a number of helpful comments.

I thank Frank Herrmann for providing the last sample binary-black-hole
evolution discussed in section~\ref{sect-sample-results/binary-BH}.
I thank Tom Goodale, Thomas Radke, and many others for their assistance
with the invaluable \program{Cactus} computational toolkit.
I thank Thomas Radke for our fruitful collaboration on \program{Cactus}
interpolators, and in particular for his work on the \program{PUGHInterp}
global interpolator.
I thank Erik Schnetter for supplying the \program{Carpet} mesh-refinement
driver and \program{CarpetInterp} global interpolator for \program{Cactus}.
I thank P.~Madderom and Tom Nicol for supplying the \program{ILUCG}
sparse-matrix subroutine.


\bibliography{references}


\end{document}


\title{A Fast Apparent-Horizon Finder for 3-Dimensional Cartesian Grids
       in Numerical Relativity: Supplemental Material}
\author{Jonathan \surname{Thornburg}}
\email{jthorn@aei.mpg.de}
\homepage{http://www.aei.mpg.de/~jthorn}
\affiliation{Max-Planck-Institut f\"ur Gravitationsphysik,
	     Albert-Einstein-Institut,
	     Am M\"uhlenberg~1, D-14476 Golm, Germany}
\preprint{AEI-2003-049}
%
%
\date{$ $Id: horizon-supplement.tex,v 1.1 2003/10/24 14:30:59 jonathan Exp $ $}


\begin{abstract}
This document contains two appendices which are omitted from the
main printed version of this paper in the interests of brevity:
appendix~\ref{app-multiprocessor-parallelization} on
``Multiprocessor and Parallelization Issues'', and
appendix~\ref{app-continuation-binary-search} on
``Searching for the Critical Parameter of a 1-Parameter Initial Data Sequence''.
These appendices also appear in the preprint-archive version of
this paper (gr-qc/0306056).
\end{abstract}


\pacs{
     04.25.Dm,	
     02.70.Bf,	
     02.60.Cb	
     }
\keywords{numerical relativity, apparent horizon, black hole}
\maketitle


\appendix
\setcounter{section}{1}


\section{Multiprocessor and Parallelization Issues}
\label{app-multiprocessor-parallelization}

\program{Cactus} (like most modern numerical relativity codes using
3-D grids), is designed to run in parallel on multiprocessor computer
systems.  \program{Cactus} uses a domain-decomposition parallelization
scheme, where each processor stores and computes the Einstein equations
on its own \defn{chunk} of the spatial grid.  Neighboring chunks overlap
slightly;
\footnote{
	 This is the 3-D Cartesian grid analog of the
	 angular interpatch ghost zones described in
	 the main printed version of this paper.
	 }
{} \program{Cactus} \defn{synchronizes} them as necessary.  An AH may
span multiple processors' grid chunks, and since an AH may move during
an evolution, in general we don't know in advance which processor(s)
those are.

Because of the domain decomposition, the multiprocessor \defn{global}
interpolator
{} used for the geometry interpolation must in general send each
interpolation point to the processor which ``owns'' that part of the
grid, do the interpolation there, and send the results back to the
requesting processor.  To ensure that every processor has a flow of
control in the interpolator code to (potentially) handle interpolation
points in its chunk of the grid, the interpolation must be a
{\em collective\/} operation: code on every processor {\em must\/}
call the interpolator synchronously (each processor's code specifying
its own choice of interpolation points).  Violations of this requirement
may result in deadlock in the interprocessor-communication code.

Taking these environmental constraints into account, I have
parallelized \program{AHFinderDirect} in the following way:
To allow the use of standard (uniprocessor) sparse matrix subroutines
for solving the Newton's-method updating routines, \program{AHFinderDirect}
assigns each AH to a single processor,
\footnote{
	 A processor may be assigned multiple AHs
	 if there are more AHs than processors.
	 }
{} and searches for that AH only on that processor.  However, if there
are multiple AHs and multiple processors, \program{AHFinderDirect}
searches for different AHs concurrently on the multiple processors.

All the processors do their Newton iterations synchronously, each
processor working sequentially through its own assigned horizon(s),
or doing dummy interpolator calls (to preserve the synchronization
across all processors) if it has no assigned horizon(s).  If/when a
processor finishes with a horizon (either locating it or failing to
locate it), the processor moves on to its next assigned horizon if
there is one, or switches to doing dummy interpolator calls if it
has no more assigned horizons left to process.
Table~\ref{tab-parallel-Newton-examples} shows two examples of this.
\footnote{
	 In part~(a) of table~\ref{tab-parallel-Newton-examples},
	 notice that after horizon~1 converges, processor~1 does
	 a dummy $\Jac$ computation before starting on the next
	 horizon.  This is slightly inefficient, but considerably
	 simplifies the algorithm by keeping the $\Theta$ and
	 $\Jac$ computations synchronized across all processors.
	 In the uniprocessor case this dummy $\Jac$ operation
	 is unnecessary, and the algorithm omits it.
	 }

\begin{table}[tbp]
%
%
\def\tspace{\qquad\qquad}	
\begin{tabular}{cc@{\qquad}cc@{}c@{\tspace}cc@{\qquad}cc@{\qquad}cc@{\qquad}cc}
\multicolumn{4}{c}{Part~(a): 2 processors}
					&& \multicolumn{8}{c}{Part~(b): 3 or more processors}		\\
\multicolumn{2}{c@{\qquad}}{\csmash{processor 1}}
	&\multicolumn{2}{c@{}}{\csmash{processor 2}}
					&& \multicolumn{2}{c@{\qquad}}{\!\csmash{processor 1}}
						&\multicolumn{2}{c@{\qquad}}{\!\csmash{processor 2}}
							&\multicolumn{2}{c@{\qquad}}{\!\csmash{processor 3}}
								&\multicolumn{2}{c}{\!\csmash{any others}}
													\\
h/it&what	&h/it&what		&&h/it &what	&h/it &what	&h/it &what	&h/it&what	\\
\cline{1-4}\cline{6-13}
1/1 &$\Theta$	&2/1 &$\Theta$		&&1/1 &$\Theta$	&2/1 &$\Theta$	&3/1 &$\Theta$	&--- &$\Theta$	\\
1/1 &$\Jac$	&2/1 &$\Jac$		&&1/1 &$\Jac$	&2/1 &$\Jac$	&3/1 &$\Jac$	&--- &$\Jac$	\\
1/2 &$\Theta$	&2/2 &$\Theta$		&&1/2 &$\Theta$	&2/2 &$\Theta$	&3/2 &$\Theta$	&--- &$\Theta$	\\
1/2 &$\Jac$	&2/2 &$\Jac$		&&1/2 &$\Jac$	&2/2 &$\Jac$	&3/2 &$\Jac$	&--- &$\Jac$	\\
1/3 &$\UU{\Theta}$
		&2/3 &$\Theta$		&&1/3 &$\UU{\Theta}$
							&2/3 &$\Theta$	&3/3 &$\Theta$	&--- &$\Theta$	\\
--- &$\Jac$	&2/3 &$\Jac$		&&--- &$\Jac$	&2/3 &$\Jac$	&3/3 &$\Jac$	&--- &$\Jac$	\\
3/1 &$\Theta$	&2/4 &$\Theta$		&&--- &$\Theta$	&2/4 &$\Theta$	&3/4 &$\UU{\Theta}$
											&--- &$\Theta$	\\
3/1 &$\Jac$	&2/4 &$\Jac$		&&--- &$\Jac$	&2/4 &$\Jac$	&--- &$\Jac$	&--- &$\Jac$	\\
3/2 &$\Theta$	&2/5 &$\UU{\Theta}$	&&--- &$\Theta$	&2/5 &$\UU{\Theta}$
									&--- &$\Theta$	&--- &$\Theta$	\\
3/2 &$\Jac$	&--- &$\Jac$	&&\multicolumn{8}{l}{}							\\
3/3 &$\Theta$	&--- &$\Theta$	&&\multicolumn{8}{l}{Abbreviations:}					\\
3/3 &$\Jac$	&--- &$\Jac$	&&\multicolumn{8}{l}{\xlsy{h/it}{what} $=$ horizon number/iteration number}
													\\
3/4 &$\UU{\Theta}$
		&--- &$\Theta$	&&\multicolumn{8}{l}{what $=$ what is this processor doing?}		
\end{tabular}
\caption[Example of Parallel Operation]
	{
	This table shows two examples of how \program{AHFinderDirect}
	finds multiple horizons in parallel in a multiprocessor
	environment, for the case where we search for 3 horizons,
	which are found (the Newton iteration converges, shown
	by the $\UU{\Theta}$) after~3, 5, and~4 iterations respectively.
	The table rows show actions at successive iterations of the
	algorithm;
	``---'' means a dummy computation (described in the main text).
	Part~(a) shows how the algorithm would work with~2 processors;
	part~(b) shows how it would work with 3~or more processors
	(the last column refers to any processors other than first~3)
	}
\label{tab-parallel-Newton-examples}
\end{table}

This algorithm requires an explicit global synchronization across all
processors at each Newton iteration:  After evaluating $\Theta$, each
processor computes a Boolean flag saying whether that processor needs to
continue iterating (this may be true for either or both of two reasons:
the Newton iteration hasn't converged yet on the current horizon, or
there is another horizon or horizons assigned to this processor which
hasn't yet been processed).  All processors then broadcast their flags,
and compute the inclusive-or of all the flags to determine whether to
continue the algorithm or exit.


\section{Searching for the Critical Parameter
	 of a 1-Parameter Initial Data Sequence}
\label{app-continuation-binary-search}

In this appendix I describe my continuation-method binary search
algorithm
for determining the ``critical'' parameter~$p_\star$ at which a common
AH appears/disappears in a 1-parameter family of initial data slices,
$p \mapsto \Sigma[p]$.  Without loss of generality I assume that
small (large) values of~$p$ do (do~not) have a common AH.

The main complication here is that \program{AHFinderDirect} needs
an initial guess for an AH shape, and if this initial guess is
inaccurate \program{AHFinderDirect} may fail to find the (an) AH.
This means that the obvious binary-search algorithm for finding $p_\ast$
isn't reliable, because a failure to find an AH doesn't rule out
the possible existence of that AH.

Instead, I use a continuation method, $p$ is ``walked up'', using
the common AHs found in smaller-$p$ slices as initial guesses for trying
to find the common AH in larger-$p$ slices.  If the algorithm fails to
find a common AH, it decreases $p$ and tries again with a smaller
``walking increment'' in $p$.
Figure~\ref{fig-continuation-binary-search-algorithm} shows this
algorithm in detail.

\begin{figure}[tbp]
\def\assign{\leftarrow}
\def\ttt{\qquad}
\begin{tabbing}
$p \assign p_\text{start}$						    \\
$\delta p \assign \delta p_\text{start}$				    \\
$G \assign G_\text{start}$				    		    \\
\ttt\=while ($|\delta p| \ge \text{tolerance}$)				  \+\\
      \{								    \\
      try to find a common AH in $\Sigma[p]$,
         using $G$ as the initial guess					    \\
      if \=(found a common AH)						  \+\\
         then \=\{							  \+\\
                $G \assign \text{the common AH just found}$		    \\
                $p \assign p + \delta p$				    \\
                \}							  \-\\
         else \>\{							  \+\\
                $\delta p \assign \thalf \delta p$			    \\
                $p \assign p - \delta p$				    \\
                \}							\-\-\\
      \}								
\end{tabbing}
\caption[Continuation-Method Binary Search Algorithm]
	{
	This figure shows the continuation-method binary
	search algorithm for finding the critical parameter
	$p$ at which a common AH appears in a 1-parameter
	family of slices.
	}
\label{fig-continuation-binary-search-algorithm}
\end{figure}
